\documentclass{qchaos}

\usepackage{amsmath,amsbsy,amssymb, amsfonts}
\usepackage{graphicx, color}
\usepackage{subfigure}
\usepackage{empheq}
\usepackage[numbers, sort&compress]{natbib}	
\usepackage[normalem]{ulem}
\usepackage{hyperref}
\usepackage[capitalise]{cleveref}
\usepackage{fancyhdr}
\usepackage{xparse}
\usepackage{mathtools}
\usepackage[shortlabels]{enumitem}

\newcommand{\Matr}[1]{\pmb{#1}}
\newcommand{\matr}[1]{#1}

\newcommand{\eps}{\epsilon}

\newcommand{\Li}{\text{\normalfont Li}}
\newcount\sbarK
\NewDocumentCommand{\sbar}{O{0} m m}{%
  \overline{#2}%
  _{%
    \!
    \sbarK=#1\relax
    \loop\ifnum\sbarK>0\relax
      \!\!\!%
      \advance\sbarK by -1\relax
    \repeat
    \lower.3ex\hbox{\(\scriptstyle #3\)}%
  }%
}
\graphicspath{{Figures/}}
\begin{document}

\title{Characterization of Chaotic Evolution in Quantum Systems Induced by Random Hermitian Matrices}
\author{Arkady Kurnosov\(^{1,2*}\)\orcid{0000-0002-0786-4392}, Sven Gnutzmann\(^3\)\orcid{0000-0002-6925-897X},   and Uzy Smilansky\(^{1}\)}

\affil{\(^1\)Department of Physics of Complex Systems, Weizmann Institute of Science,
Rehovot 76100, Israel}

\affil{\(^2\)Department of Physics, Faculty of Sciences, Holon Institute of Technology,
Holon 5810201, Israel}

\affil{\(^3\)School of Mathematical Sciences, University of Nottingham, University Park,
Nottingham NG7 2RD, United Kingdom}

\affil{\(^*\)Author to whom any correspondence should be addressed.}

\email{akurnoso@tulane.edu}


\begin{abstract}
In a recent paper \cite{gnutzmann2024}, a semiclassical Lyapunov exponent
associated with a quantum Hamiltonian represented by a finite-dimensional
Hermitian matrix was defined and placed on a mathematical foundation.
The Lyapunov exponent characterizes the early stages of the evolution toward
the ergodic state, while the late stages are characterized by the spectral gap
of the corresponding Markov matrix. Here, we apply this formalism to five random-matrix ensembles.
For each ensemble, we derive the mean Lyapunov exponent, its variance, and the
spectral gap as functions of energy.
We also present the corresponding thermal averages.
Extensive numerical data are compared with the theoretical predictions.
\end{abstract}
\section*{Introduction}
The main signature of classical deterministic chaos is the
sensitive dependence of the dynamics on initial conditions.
It is characterized by a Lyapunov exponent (LE), denoted by \(\Lambda\), which is a non-negative quantity such that the phase-space distance
between nearby trajectories increases exponentially as \(e^{\Lambda t}\)
\cite{ott2002,cvitanovic2020}.
The concept of an LE has been generalized to stochastic
evolution, where
it describes the exponential divergence of typical stochastic trajectories \cite{barra2001,laffargue2016}.
In this context, one speaks of stochastic chaos if the stochastic LE is positive.
Both deterministic and stochastic LEs are rigorously defined and discussed 
in the mathematical theories of dynamical and ergodic systems.

These rigorous definitions cannot be transferred directly to quantum mechanics,
as the linearity of quantum dynamics leaves no room for a positive LE.
The search for ``fingerprints'' of classically chaotic dynamics in their quantized
analogues has fueled a decades-long discussion about the definition of
``quantum chaos''.
Many quantum systems have an underlying classical Hamiltonian dynamics that is
recovered in the semiclassical limit \(\hbar\to0\).
These efforts have yielded numerous results on the quantum signatures of
classical chaos and integrability in the semiclassical regime
\cite{gutzwiller1990,haake2018}.
The Bohigas--Giannoni--Schmit conjecture \cite{bohigas1984} predicts the spectral
statistics, while the Shnirelman theorem concerns the distribution of the
eigenvectors \cite{shnirelman1974}.
However, this approach is not applicable to a large class of  quantum
systems that do not possess a classical limit, such as, e.g., systems of particles with spin 1/2 for which there is no classical analogue.

While there is no rigorous definition of an LE in quantum mechanics, it has
long been known that some quantum measures exhibit exponential
growth over a certain time scale.
One example is the Loschmidt echo \cite{goussev2012}, which measures the
divergence between the evolution of a quantum state under a given Hamiltonian
and that of a state evolving under a perturbed Hamiltonian.

More recently, the notion of a quantum Lyapunov exponent has been introduced
\cite{shenker2014}, based on earlier work on quantum information scrambling in
black holes \cite{hayden2007,sekino2008}.
It describes the exponential rate of quantum information scrambling, i.e.,
the process by which local information becomes distributed throughout Hilbert
space \cite{landsman2019,mi2021,harris2022,garcia2023}.
In many-body systems, it describes how a subsystem consisting of a few particles
becomes entangled with the entire system through quantum evolution.
It can be quantified through the exponential growth of out-of-time-ordered
correlators \cite{larkin1969,garcia-mata2023}.

The existence of a temperature-dependent upper bound on quantum (OTOC) LEs
\cite{maldacena2016} has led to the notion of perfect scramblers, such as SYK
systems \cite{kitaev2015}.
It is generally believed, and has been shown in special cases, that the quantum
LE coincides with the LE of an underlying classical dynamics.

In the previous work \cite{gnutzmann2020,gnutzmann2024}, an alternative definition
of the quantum LE was considered, based on the LE of a suitably defined
underlying classical Hamiltonian flow.
Using graph-theoretic ideas inspired by the theory of quantum graphs, the notion
of a quantum Poincar\'e map \(\Matr{U}(E)\) was introduced.
This is an energy-dependent unitary map that encodes the dynamics of a given
quantum system in a fixed observational basis.
Its matrix elements are quantum transition amplitudes on a discrete ``phase
space.''
The evolution is described by successive applications of \(\Matr{U}\), and the
(topological) time is the number of steps.
The fact that the time variable is an integer, rather than a physical measure
of time, allows the energy variable to be used without conflicting with the
energy--time uncertainty relation.

Taking the squared moduli of the quantum transition amplitudes yields the corresponding
transition probabilities, which define a classical stochastic Markov process
that depends on the energy.
It was suggested in Ref.~\cite{gnutzmann2024} that the LEs of this Markov process
may serve as indicators of quantum chaos and quantum information scrambling and
can be calculated numerically without large-scale simulations.
More specifically, it was shown that the behavior of the mean LEs, and
especially of certain local variants, distinguishes between localized and
extended (ergodic) wavefunctions.
Indications were also given of how these quantities enter dynamical observables
such as out-of-time-ordered correlators. 

In the present work, we continue the characterization of the LEs of the
corresponding Markov process for typical quantum-chaotic systems by considering
finite-dimensional Hamiltonians drawn from five ensembles: the ensemble of
adjacency matrices of \(d\)-regular graphs with a given number of vertices;
the Gaussian orthogonal ensemble (GOE) and Gaussian unitary ensemble (GUE) of
random-matrix theory; and the corresponding tridiagonal G\(\beta\)E ensembles
\cite{Dumitriu2002} with \(\beta=1,2\).
For each ensemble, we compute the mean LE, its variance, and the spectral gap
of the corresponding stochastic process. Our results are due to both numerical simulations and theoretical computations.

The paper is organized as follows.
In \cref{Sec:Prelim}, we provide the necessary details of our method and state
the resulting expressions for the mean Lyapunov exponent and its variance.
It is important to note that, for a single matrix, the LE is defined by averaging
over all trajectories consistent with the Markov process determined by that
matrix.
We also show that the LE provides a measure of the rate of information loss
during the early stages of the evolution.

In \cref{Sec:MeanLE}, we study the mean LE for the five ensembles.
The main conclusion of this section is that the LE is a self-averaging quantity:
after appropriate scaling, it converges to its ensemble mean.
We also show that, for a class of large sparse matrices, the mean LE depends only
on the degree, i.e., the number of nonzero off-diagonal elements in each row,
and not on the matrix dimension, as occurs, for example, in certain interacting
spin models.

In \cref{Sec:Variance}, we present the results for the ensemble-averaged LE
variance.
In general, the variance is not self-averaging, and we therefore present its
probability density function (PDF).
The PDF exhibits a power-law decay whose exponent depends critically on the
ensemble.
The study of the variance requires knowledge of the spectral properties of the
corresponding stochastic matrices.
We therefore present their spectra together with the distribution of the
spectral gap between the Perron--Frobenius eigenvalue and the eigenvalue of
second-largest modulus.
The spectral gap measures the rate at which the system approaches equilibrium,
thus complementing the information provided by the LE.

In \cref{Sec:ThermalAverage}, we study the temperature dependence of the LE,
obtained from the energy dependence derived in the preceding sections.
Finally, we summarize and discuss the main results and outline directions for
future work.

\section{Preliminaries}\label{Sec:Prelim}
\subsection{The quantum Poincar\'e map}\label{Sec:PrelimQuantum}

We consider an arbitrary finite-dimensional quantum system in a fixed observational basis,
such that its Hamiltonian is represented by a \(V\times V\) Hermitian matrix \(\Matr{H}\).
Following previous work \cite{smilansky2007,gnutzmann2020,gnutzmann2024}, we associate
\(\Matr{H}\) with an underlying graph \(\mathcal{G}\) consisting of \(V\) vertices and
having the adjacency matrix
\begin{equation}
A_{vw} =
\begin{cases}
1 & \text{if \(v\neq w\) and \(H_{vw}\neq 0\),}\\
0 & \text{otherwise.}
\end{cases}
\end{equation}
The degree of vertex \(v\) is denoted by \(d_v=\sum_w A_{vw}\).
Two vertices \(v\) and \(w\) are connected by an edge if \(A_{vw}=1\).
With each such edge, we associate two directed edges, denoted by \((vw)\)
(directed from \(w\) to \(v\)) and \((wv)\) (directed in the opposite direction).
The total number of directed edges is \(D=\sum_{v=1}^{V}d_v=\sum_{v,w=1}^{V}A_{vw}\), 
which is equal to the sum of all elements of the adjacency matrix.

The vertex set \(\mathcal{V}\) of the graph serves as the configuration space for the
dynamics, while the corresponding phase space \(\mathcal{D}\) consists of all directed
edges.
For a given directed edge \(e=(vw)\), its origin is \(o(e)=w\), and its terminus is
\(\tau(e)=v\).
Classical trajectories in phase space are sequences of connected directed edges \(e_k\)
satisfying\(o(e_{k+1})=\tau(e_k)\).
For any directed edge \(e=(vw)\), we denote the same edge with the reverse orientation
by \(\hat e=(wv)\).
We use a term ``phase space'' for \(\mathcal{D}\) because it contains information on 
the present and the next vertices, similar to the role played by coordinates and 
momenta in classical dynamics. It is convenient for each vertex \(v\) to define its 
incoming and outgoing stars \(\mathcal{S}^{\mathrm{in}}_{v}\) 
and \(\mathcal{S}^{\mathrm{out}}_{v}\), by the set of directed edges with \(\tau(e) = v\) or 
\(o(e) = v\), respectively.

One may relate the dynamics in the configuration space of vertices to unitary dynamics
on the phase space of directed edges of the underlying graph by considering the propagator
matrix (also known as the resolvent or Green matrix)
\begin{equation}
\Matr{W}(E)=\left(\Matr{H}-E\Matr{I}_{V}\right)^{-1},
\end{equation}
where \(E\) is a complex spectral parameter lying outside the spectrum of \(\Matr{H}\).
In the following, we refer to \(E\) as the energy and assume that it is real unless
explicitly stated otherwise.
The connection to the phase-space dynamics is established by the identity
\begin{equation}\label{Eq:resolvent_identity}
\Matr{W}(E)=\Matr{M}(E)
-2\,\Matr{L}(E)
\left[\Matr{I}_{D}-\Matr{U}(E)\right]^{-1}\Matr{R}(E),
\end{equation}
where the right-hand side contains five matrices, which we define and explain below.

To write the matrix \(\Matr{U}\) explicitly, we adopt the following
convention. Square roots \(\sqrt{H_{vw}}\) of off-diagonal matrix elements
(\(v\neq w\)) will appear frequently. We choose the branch with
non-negative real part and require \(\sqrt{H_{vw}}=\bigl(\sqrt{H_{wv}}\bigr)^{\ast}\).
If \(H_{vw}=-x^2\), where \(x>0\) and \(v>w\), then \(\sqrt{H_{wv}}=ix=-\sqrt{H_{vw}}\).

Let us also introduce the Gershgorin radii
\begin{equation}
\Gamma_v=\sum_{w\neq v}\left|H_{vw}\right|
\end{equation}
for each vertex \(v\).
We can now define all the matrices appearing on the right-hand side of
\cref{Eq:resolvent_identity} and explain their physical significance.
The diagonal \(V\times V\) matrix \(\Matr{M}(E)\) is given by
\begin{equation}
M_{vw}(E)=\delta_{vw}\frac{1}{H_{vv}-\mathrm{i}\Gamma_v-E},
\end{equation}
and describes local decay on a time scale \(\tau_v\propto\hbar/\Gamma_v\).

Next, the \(D\times V\) matrix \(\Matr{R}(E)\) is defined by
\begin{equation*}
R_{ev}(E)=
\delta_{o(e)v}
\frac{\sqrt{H_{\tau(e)v}}}{H_{vv}-\mathrm{i}\Gamma_v-E}.
\end{equation*}
Thus, \(\Matr{R}(E)\) maps a quantum amplitude at a given vertex \(v\)
in the original Hilbert space to amplitudes on the directed edges in the outgoing
star of \(v\).
Similarly, the \(V\times D\) matrix \(\Matr{L}(E)\), with matrix elements
\[
L_{ve}(E)=
\delta_{v\tau(e)}
\frac{\sqrt{H_{vo(e)}}}{H_{vv}-\mathrm{i}\Gamma_v-E},
\]
maps a set of amplitudes on the directed edges in the incoming star of \(v\)
to a single amplitude at the vertex.

The matrices \(\Matr{R}(E)\) and \(\Matr{L}(E)\) provide a direct connection
between the configuration space of vertices and the phase space of directed edges.
Note that their matrix elements contain the same resonance denominators as those of
\(\Matr{M}(E)\).

Finally, we come to the \emph{quantum Poincar\'e map} \(\Matr{U}(E)\),
which will play a central role in the remainder of this manuscript.
It is a unitary \(D\times D\) matrix defined by
\begin{equation}\label{Eq:MatrU}
U(E)_{e'e}
=
\delta_{o(e')\tau(e)}
\sigma_{\tau(e')o(e)}^{(\tau(e))}(E)
\end{equation}
in terms of the \(d_v\times d_v\) unitary vertex-scattering matrices
\begin{equation}\label{Eq:SigmaScatter}
\sigma_{w'w}^{(v)}(E)
=
\mathrm{i}\delta_{w'w}
-2\frac{\sqrt{H_{w'v}H_{vw}}}
{H_{vv}-\mathrm{i}\Gamma_v-E},
\end{equation}
where \(w\) and \(w'\) run over all vertices adjacent to \(v\).

For real energies \(E\), the quantum Poincar\'e map defines unitary dynamics
on the directed edges consistent with the connectivity of the underlying graph:
a nonzero transition amplitude from a directed edge \(e\) to a directed edge \(e'\)
exists only if the terminus of \(e\) is the origin of \(e'\).
The right-hand side of \cref{Eq:resolvent_identity} contains the term
\begin{equation}
\left[\Matr{I}_{D}-\Matr{U}(E)\right]^{-1}
=
\lim_{\epsilon\to0^+}
\sum_{n=0}^{\infty}e^{-\epsilon n}\Matr{U}(E)^n,
\end{equation}
which shows that the propagator can be expanded as a sum over trajectories
on the underlying graph. The amplitude associated with each trajectory contains
the product of the corresponding transition amplitudes of the quantum Poincar\'e map.
Here, \(\Matr{S}(E)\) represents the contribution of ``zero-length'' trajectories
that never enter a directed edge.

The identity \cref{Eq:resolvent_identity} is a direct consequence of the
determinant identity
\begin{equation}
\det\left[E\Matr{I}_{V}-\Matr{H}\right]
=
\frac{
\prod\limits_{v=1}^V\left(H_{vv}-\mathrm{i}\Gamma_v-E\right)}
{2^{D/2}}
\det\left[\Matr{I}_{D}-\Matr{U}(E)\right],
\end{equation}
proved in \cite{gnutzmann2020}.
Equation \cref{Eq:resolvent_identity} follows by taking the logarithmic derivative
of this identity with respect to the matrix elements of the Hamiltonian.

The quantum Poincar\'e map may be written as
\begin{equation}\label{Eq:UPSigma}
\Matr{U}(E)=\Matr{P}\Matr{\Sigma}(E),
\end{equation}
where \(\Matr{\Sigma}(E)\) is a unitary block-diagonal matrix with \(V\)
diagonal blocks \(\Matr{\sigma}^{(v)}(E)\).
The matrix \(\Matr{P}\) is a unitary permutation matrix that maps each directed
edge \(e\) to its reverse \(\hat e\); that is,
\(P_{e'e}=\delta_{e'\hat e}\). 

\subsection{The classical Poincar\'e map and the Lyapunov exponent}
\label{Sec:PrelimClassic}

Discrete classical dynamics is obtained by constructing the Poincar\'e--Markov map
\(\Matr{B}(E)\) from \(\Matr{U}(E)\), replacing the quantum transition
\emph{amplitudes} with the corresponding transition \emph{probabilities}
\cite{kottos1997}:
\begin{equation}
B_{e'e}(E)=\left|U_{e'e}(E)\right|^2.
\end{equation}
The matrix \(\Matr{B}(E)\) is bistochastic:
\(\sum_{e\in\mathcal{D}}B_{e'e}(E)
=\sum_{e'\in\mathcal{D}}B_{e'e}(E)=1\).
Transitions \(e\mapsto e\) are impossible; therefore, \(B_{ee}(E)\equiv0\).
Analogously to \cref{Eq:UPSigma}, the matrix \(\Matr{B}(E)\) can be written as
\begin{equation}\label{Eq:BPX}
\Matr{B}(E)=\Matr{P}\Matr{\Pi}(E),
\end{equation}
where \(\Matr{\Pi}(E)\) is a block-diagonal Hermitian matrix with elements
\(\Pi_{e'e}(E)=|\Sigma_{e'e}(E)|^2\).

The probability \(p_e(t)\) of being on the directed edge \(e\) after \(t\)
time steps evolves according to
\[
p_{e'}(t+1)=\sum_{e\in\mathcal{D}}B_{e'e}(E)p_e(t).
\]
The spectrum of \(\Matr{B}(E)\) lies within the unit disk in the complex plane.
The matrix \(\Matr{B}(E)\) is pseudo-Hermitian, since
\(\Matr{B}^\dagger
=\Matr{P}\Matr{B}\Matr{P}^{-1}\).
Therefore, its eigenvalues are either real or occur in complex-conjugate pairs
\cite{mostafazdeh2010}.
Moreover, the spectrum always contains the eigenvalue \(\nu_0=1\), corresponding
to the Frobenius eigenvector \(|0)\propto(1,\dots,1)^\top\).
If all eigenvalues other than \(\nu_0\) lie strictly inside the unit disk,
the evolution is mixing, and the probability distribution converges exponentially
to the uniform distribution.
The spectral gap
\(\min_{k\neq0}(1-|\nu_k|)\)
determines the asymptotic rate of convergence to equilibrium.
Moreover, the spectrum of \(\Matr{B}(E)\) enters the expression for the variance
of the Lyapunov exponent.

In the present context, a trajectory of \(t\) time steps is a sequence \(\xi_t=(e_j)_{j=0}^{t}\) of connected directed edges starting at a prescribed directed edge \(e_0\). The probability to remain on the chosen  trajectory is:
\begin{equation}
\mathcal{P}(\xi_{t}) = \prod\limits_{j=1}^{t}B_{e_{j} e_{j-1}}, \quad o(e_{j}) = \tau(e_{j-1}).
\end{equation}

Because of the number of directed edges, \(D\), is finite, the corresponding dynamics is a {\it shift of finite type}, and the probability to remain on the prescribed trajectory of length \(t\gg D\), decays exponentially. Therefore, the mean value of the Lyapunov exponent  is defined as:
\begin{equation}\label{Eq:MeanLE-Def}
\langle\Lambda\rangle = -\lim_{t\to\infty}\frac{1}{t}\left\langle \ln\left[\mathcal{P}(\xi_{t})\right]\right\rangle_{\xi_{t}}, 
\end{equation} 
where the average is over all trajectories of length \(t\) and all initial directed edges \(e_{0}\). Since the trajectories are labeled by a finite number \(D\) of indices one can replace the infinite sum \cref{Eq:MeanLE-Def} by 
\begin{equation}\label{Eq:MeanLE-Th0}
\left\langle\Lambda(E)\right\rangle =  -\frac{1}{D}\sum\limits_{e, e^{\prime}}B_{e^{\prime}e}(E)\ln B_{e^{\prime}e}(E),
\end{equation}
where \(B_{e^{\prime}e}(E)\) is the probability of making a transition from \(e\) to \(e^{\prime}\).

\begin{figure}[ht]
 \begin{center}
        \includegraphics[width=.9\textwidth]{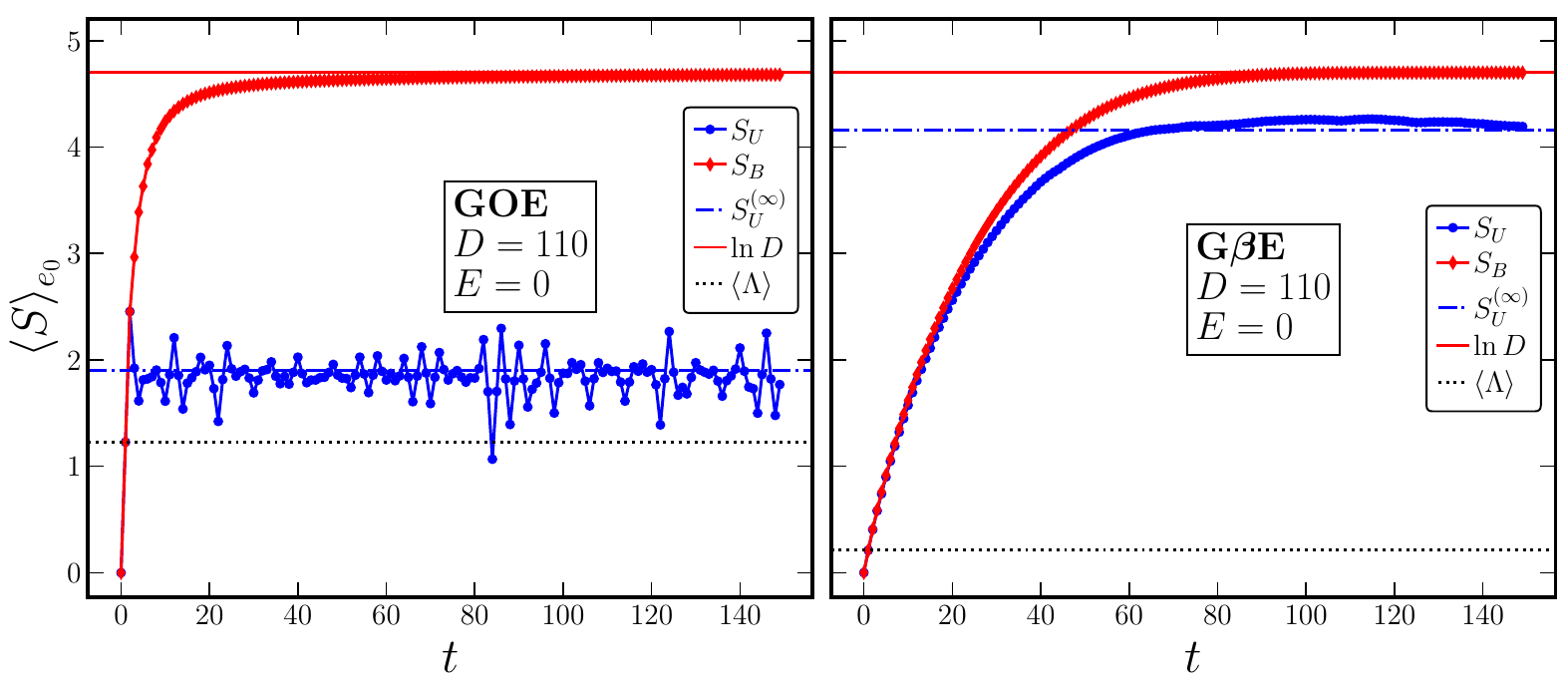}
 \caption{Entropy as a function of topological time for a single realization,
averaged over all localized initial states. Left: GOE. Right: G\(\beta\)E with
\(\beta=1\).}
\label{Fig:Entropy_vs_t}
\end{center}
\end{figure}

Consider the entropy production from an initial state localized on the directed
edge \(e_0\).
Under quantum evolution, the entropy after \(t\) time steps is given by
\[
S_U(t)=-\sum\limits_n
\left|(\Matr{U}^t)_{e_n,e_0}\right|^2
\ln\left[\left|(\Matr{U}^t)_{e_n,e_0}\right|^2\right].
\]
For \(t=1,2\), these expressions coincide with their semiclassical counterparts,
\(S_B(t)\).
Averaging over all \(D\) localized initial states, one obtains
\(\langle S_U(1)\rangle_{e_0} = \langle S_B(1)\rangle_{e_0} = \langle\Lambda\rangle\),
where \(\langle\Lambda\rangle\) is the mean LE defined above.
Thus, the mean LE characterizes the initial rate of entropy production and
provides information about the early stages of relaxation toward equilibrium.

For large \(t\), the quantum entropy \(S_U(t)\) fluctuates around a saturation
value \(S_U^{(\infty)}\), which is lower than the stochastic entropy of the
equilibrium state, \(\ln D\).
This behavior is illustrated in \cref{Fig:Entropy_vs_t}, which shows numerical
results for \(S_U(t)\), together with its semiclassical counterpart \(S_B(t)\)
and the estimate of \(S_U^{(\infty)}\) derived in \cref{Sec:Sinf}.
A complete theory of entropy evolution will be presented in a subsequent paper.

Ergodic theory formalism \cite{parry1990zeta,barra2001,gnutzmann2024} provides a powerful method to compute the mean of the LE and its higher moments. This is done by introducing an auxiliary matrix \(\Matr{Q}(E, \varepsilon)\): \(Q_{e^{\prime}e}(E, \varepsilon) = \left[B_{e^{\prime} e}(E)\right]^{1 + \varepsilon}\). Denoting the eigenvalue of \(\Matr{Q}(E,\varepsilon)\) with the largest real
part by \(\mu(\varepsilon)\), the mean LE and its variance are given by
\begin{equation}\label{Eq:MeanVar}
\langle\Lambda\rangle = -\frac{\partial\ln\mu}{\partial\varepsilon}\Big|_{\varepsilon = 0}, \quad
\langle(\delta\Lambda)^{2}\rangle = \frac{\partial^{2}\ln\mu}{\partial\varepsilon^{2}}\Bigr|_{\varepsilon = 0}.
\end{equation}
One can rewrite
\[
Q_{e^{\prime} e}(E, \varepsilon) =
 B_{e^{\prime} e} + \varepsilon B_{e^{\prime} e}\ln B_{e^{\prime} e} + \frac{\varepsilon^{2}}{2}B_{e^{\prime} e}\left[\ln B_{e^{\prime} e}\right]^{2} + \mathit{o} [\varepsilon^{2}],
\]
or, introducing matrices \(G_{e^{\prime} e} = B_{e^{\prime} e}\ln B_{e^{\prime} e}\), and \(F_{e^{\prime} e} = B_{e^{\prime} e}\left[\ln B_{e^{\prime} e}\right]^{2}\), 
\begin{equation}\label{Eq:PerturbMatrix}
\Matr{Q}(E, \varepsilon) \approx \Matr{B} + \varepsilon\Matr{G} + \frac{\varepsilon^{2}}{2}\Matr{F}. 
\end{equation}
The largest eigenvalue, therefore, can be written as  \(\mu(\varepsilon) = 1 + \varepsilon\mu^{(1)} + (\varepsilon^{2}/2)\mu^{(2)} + \dots\) within a framework of perturbation theory for non-Hermitian.
Using the explicit definition of the perturbation from \cref{Eq:PerturbMatrix}, and collecting terms of the same order with respect to \(\varepsilon\), we obtain the moments of the LE distribution from \cref{Eq:MeanVar}:
\begin{equation}\label{Eq:MeanLE-Th}
\left\langle\Lambda(E)\right\rangle =  -\frac{1}{D}\sum\limits_{e, e^{\prime}}B_{e^{\prime} e}(E)\ln B_{e^{\prime} e}(E)
\end{equation}

\begin{equation}\label{Eq:LambdaMoments}
\begin{gathered}
\langle\left[\delta\Lambda(E)\right]^2\rangle = \langle\left[\delta\Lambda(E)\right]^2\rangle_F + \langle\left[\delta\Lambda(E)\right]^2\rangle_G -\langle\Lambda\rangle^2,\\
\langle(\delta\Lambda)^2\rangle_F = \frac{1}{D}\sum\limits_{e,e'}B_{e'e}\left[\ln B_{e'e}\right]^2,
\quad
\left\langle(\delta\Lambda)^2\right\rangle_G = 2\sum\limits_{k\neq0}\frac{(0|\Matr{G}|r_k)(l_k|\Matr{G}|0)}{(l_k|r_k)\left(1-\nu_k\right)}.
\end{gathered}
\end{equation}
Here, \(|r_k)\), \((l_k|\), and \(\nu_k\) are the right eigenvectors, left
eigenvectors, and eigenvalues of \(\Matr{B}\), respectively, and \(|0)\) is the
Frobenius vector defined above.
We use \((l_k|=|l_k)^\top\), where \(\top\) denotes a simple transpose rather
than a Hermitian conjugate.

For simplicity, we refer to
\(\langle(\delta\Lambda)^2\rangle_F\) and
\(\langle(\delta\Lambda)^2\rangle_G\) as the \(F\)- and \(G\)-components of the
variance, respectively.
As we will see, the relative contributions of the two components and the centering term, \(\langle\Lambda\rangle^{2}\), vary
considerably among the different models.
Note that the result for \(\langle\Lambda\rangle\) in
\cref{Eq:MeanLE-Th} is identical to \cref{Eq:MeanLE-Th0}, which was obtained
from a different perspective.

A detailed discussion of the spectral properties of \(\Matr{B}\) is beyond the
scope of the present study and will be reported in a future publication
\cite{KurnosovSmilansky2026}.
Here, we mention only the possibility that a pseudo-Hermitian matrix may exhibit
an exceptional point of second order (EP2).
An EP2 is a special type of degeneracy occurring at a parameter value
\(E=E_{\mathrm{ep}}\), at which two eigenvalues become degenerate and their
corresponding eigenvectors coalesce.
Moreover, these eigenvectors become \textit{self-orthogonal}, i.e.,
\(c_k=(L_k|R_k)\to0\) continuously as \(E\to E_{\mathrm{ep}}\).

The expression for the \(G\)-component in \cref{Eq:LambdaMoments} nevertheless
remains applicable because
\(\langle(\delta\Lambda)^2\rangle_G\) has a removable singularity at
\(E=E_{\mathrm{ep}}\).
Consequently, the variance of the LE does not become singular at possible
degeneracies in the spectrum of \(\Matr{B}\).
This can be seen by expanding the eigenvalues and eigenvectors in
\textit{Puiseux series} in the vicinity of the EP2.


\section{Applications to random matrix ensembles}\label{Sec:Applications}
In the previous sections, we introduced several quantities associated with a
single matrix \(\Matr{H}\), including the mean LE
\(\langle\Lambda\rangle\), its variance
\(\langle(\delta\Lambda)^2\rangle\), and the spectral gap.
Here, the averages are taken over the trajectories induced by that matrix.
However, \(\Matr{H}\) may itself be a single realization drawn from an ensemble
of random matrices \(\{\Matr{H}_i\}\).
Each realization therefore produces a value \(X_i\) of a given observable \(X\),
and the ensemble defines a corresponding distribution \(p_X(x)\).
We use an overbar, \(\overline{\,\cdot\,}\), to denote an ensemble average and to distinguish it from the
trajectory average for a single matrix realization.
In numerical calculations, the same notation denotes the corresponding
finite-sample estimate:
\begin{equation}\label{Eq:averageVSmean}
\sbar{X}{}
=
\mathbb{E}_{\Matr{H}}[X]
\approx
\frac{1}{N}\sum_{i=1}^{N}X_i,
\end{equation}
where \(N\) is the number of matrix realizations in the numerical sample.
This notation will be used for the mean LE, the LE variance, the spectral gap,
and their higher moments.

In the present section, we consider five matrix ensembles: adjacency matrices of
\(d\)-regular graphs, the Wigner--Dyson ensembles GOE and GUE, and the corresponding
G\(\beta\)E ensembles with \(\beta=1,2\).
The G\(\beta\)E ensembles consist of tridiagonal random matrices introduced in
Ref.~\cite{Dumitriu2002}.
For \(\beta=1,2\), the spectra of G\(\beta\)E have exactly the same spectral
statistics as those of GOE and GUE, respectively.
Their eigenvector behavior, however, is different and exhibits a transition from
localized to delocalized states for \(\beta>2\) \cite{breuer2007}.
The distributions of the mean LEs and the LE variances clearly demonstrate that the
present method is sensitive not only to the spectrum but also to the matrix
representation.

The results for the five ensembles are presented in the following two sections.
The first is devoted to a comparison of the mean LEs, and the second to the LE
variances and the spectral gaps.
To compare the results on an equal footing, we normalize the Hamiltonians of all
the models so that the support of their asymptotic spectral distributions is
the interval \([-1,1]\). The normalization constant will be denoted by \(\xi\).

\subsection{Mean Lyapunov exponent}\label{Sec:MeanLE}
\subsubsection{\(d\)-regular graphs}\label{Sec:MeanLE-dreg}

For a \(d\)-regular graph, the chaotic dynamics is generated by the Hamiltonian
\(\Matr{H}^{(\mathrm{d})}=\xi\Matr{A}^{(\mathrm{d})}\),
where \(\Matr{A}^{(\mathrm{d})}\) is the adjacency matrix of the graph.
The normalization parameter \(\xi=1/\sqrt{4(d-1)}\) is chosen so that, in the
limit of large graph size, the nontrivial eigenvalues of
\(\Matr{H}^{(\mathrm{d})}\) lie in the interval \([-1,1]\), while the exceptional
eigenvalue is equal to \(\xi d\).
The elements of the Poincar\'e--Markov map \(\Matr{B}\) are given by
\begin{equation}\label{Eq:Bd}
B_{e'e}(E) = \left\{
\begin{aligned}
&p_r = \frac{4}{d^2\Delta^2}, &e'=\hat e,\\
&p_t = 1-\frac{4(d-1)}{d^2\Delta^2}, &o(e')=\tau(e)\ \text{and}\ e'\neq\hat e,\\
&0,   & \text{otherwise},
\end{aligned}
\right.
\quad
\Delta^2=1+\left(\frac{E}{\xi d}\right)^2.
\end{equation}
By construction, \(\Matr{B}\) is a \(D\times D\) matrix, with \(D=Vd\).
At each vertex, there is one reflection probability \(p_r\) and \(d-1\)
identical transmission probabilities \(p_t\).
Thus, the sum of each column and each row is \(p_r+(d-1)p_t=1\).
Substituting \cref{Eq:Bd} into \cref{Eq:MeanLE-Th}, we obtain
\begin{equation}\label{Eq:LambdaDreg}
\langle\Lambda_{\mathrm{d}}\rangle
=
\frac{4(d-1)}{d^2+\lambda^2}
\ln\left[\frac{d^2+\lambda^2}{4}\right]
+
\frac{(d-2)^2+\lambda^2}{d^2+\lambda^2}
\ln\left[
\frac{d^2+\lambda^2}{(d-2)^2+\lambda^2}
\right],
\quad
\text{where}
\qquad
\lambda=\frac{E}{\xi}.
\end{equation}
For this ensemble, the mean LE does not depend on the number of vertices \(V\).
Moreover, the trajectory average is the same for every realization of the random
matrix \(\Matr{A}^{(\mathrm{d})}\), so that
\(\sbar{\Lambda}{\mathrm{d}}
\equiv\langle\Lambda_{\mathrm{d}}\rangle\).
This is because the contributions to the LE for a \(d\)-regular graph are
independent of the vertex at which the transition occurs.

\begin{figure}[ht]
 \begin{center}
        \includegraphics[width=.5\textwidth]{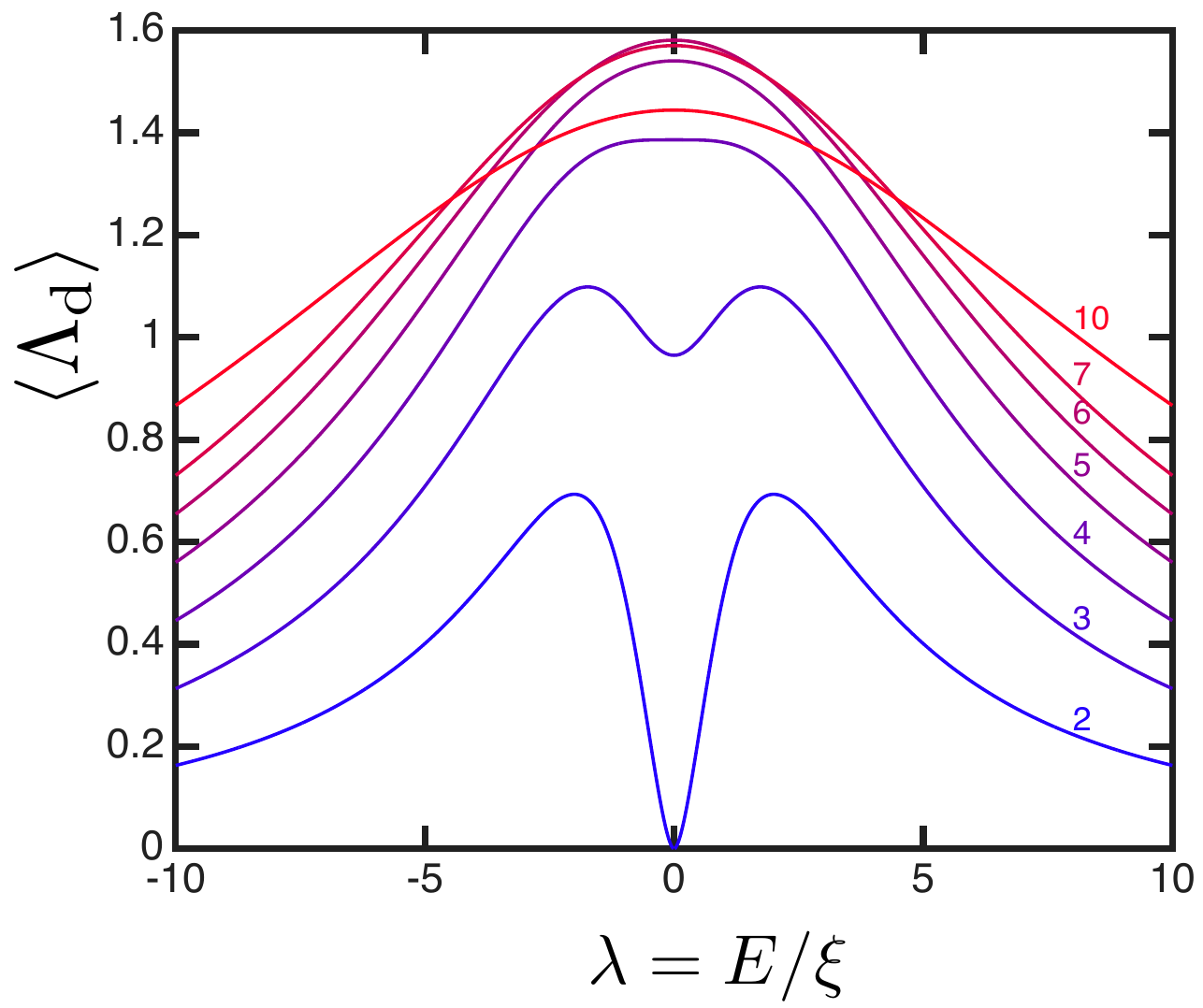}
 \caption{
Mean LE of the \(d\)-regular graph as a function of
\(\lambda=E/\xi\), see \cref{Eq:LambdaDreg}, for different values of \(d\).
The dependence on the graph connectivity is non-monotonic, with the maximum of
\(\langle\Lambda_{\mathrm d}\rangle\) attained at \(d=6\).
}
\label{Fig:Dreg_LE}
\end{center}
\end{figure}

The dependence of the mean LE on the connectivity \(d\) is illustrated in
\cref{Fig:Dreg_LE}.
It stops increasing and begins to decrease between \(d=6\) and \(d=7\),
coinciding with a structural change: the reflection probability becomes
\(p_r>1/2\).
For \(d\gg1\), the reflection probability,
\(p_r=1-4/d+\mathcal{O}(d^{-2})\), approaches \(1\).
The dominant trajectories in this limit are backscattering trajectories, which
retrace their paths rather than spreading over the graph.
Therefore, the mean LE approaches zero.

For \(d\gg1\), \cref{Eq:LambdaDreg} becomes
\begin{equation}\label{Eq:LambdaDregApprox}
\sbar{\Lambda}{\mathrm{d}}(E)
\xrightarrow{d\gg1}
\frac{4\ln\left(\frac{d^2\Delta^2}{4}\right)}{d\Delta^2}
+\frac{4}{d\Delta^2}
+o\left(d^{-1}\right).
\end{equation}
We will use this expression to analyze the asymptotic behavior of the GOE and
GUE ensembles.
 
\subsubsection{Wigner--Dyson ensembles}\label{Sec:MLE-GOEGUE}

In this section, we focus on the GOE and GUE ensembles.
The GOE consists of real symmetric \(V\times V\) matrices whose independent
entries are normally distributed: the diagonal entries follow
\(\mathcal{N}[0,1/(2V)]\), while the off-diagonal entries follow
\(\mathcal{N}[0,1/(4V)]\).
In the GUE, the diagonal entries follow \(\mathcal{N}[0,1/(4V)]\), while the
off-diagonal entries are complex, with independent real and imaginary parts
distributed according to \(\mathcal{N}[0,1/(8V)]\).
The corresponding graph is fully connected, with \(d=V-1\).
For large matrices \(d\approx V\gg1\), there is no need to distinguish between these
two parameters.
We therefore mostly use \(d\) to facilitate comparison between the Wigner--Dyson
ensembles and \(d\)-regular graphs of the same degree. 

We use the following arguments to derive an analytical expression for the mean LE.
First, by the central limit theorem, the Gershgorin radii concentrate around
their mean:
\(\Gamma_v=\xi d\left(1+\mathcal{O}(d^{-1/2})\right)\), where
\(\xi_1=1/\sqrt{2\pi d}\) for the GOE and
\(\xi_2=\sqrt{\pi/(16d)}\) for the GUE.

Second, we compute the ensemble average of each term in the sum in
\cref{Eq:MeanLE-Th}.
Retaining only terms of order \(\ln(d)/d\) and \(1/d\), the ensemble-averaged
mean LE can be expressed as
\begin{equation}\label{Eq:LambdaGOEGUE}
\overline{\Lambda}(E)
=
\left\{
\begin{aligned}
&\frac{4}{\Delta^2}
 \frac{\ln\left[\frac{d^2\Delta^2}{4}\right]}{d}
+\frac{4}{\Delta^2}
 \frac{\gamma-\ln\pi+1}{d},
&&
\Delta^2=1+\left(\frac{E}{\xi_1d}\right)^2,
\quad \text{GOE},
\\
&\frac{4}{\Delta^2}
 \frac{\ln\left[\frac{d^2\Delta^2}{4}\right]}{d}
+\frac{4}{\Delta^2}
 \frac{\gamma+\ln\pi-1}{d},
&&
\Delta^2=1+\left(\frac{E}{\xi_2d}\right)^2,
\quad \text{GUE},
\end{aligned}
\right.
\end{equation}
where \(\gamma\approx0.577\) is the Euler--Mascheroni constant.

Comparison of \cref{Eq:LambdaGOEGUE} with \cref{Eq:LambdaDregApprox} shows that
the mean LE for the Wigner--Dyson ensembles can be approximated by that generated
by a \(d\)-regular graph.
Interestingly, at \(E=0\), the differences among the three models appear only
in the \(1/d\) term, so that
\(\sbar{\Lambda}{\mathrm{GOE}}
<\sbar{\Lambda}{\mathrm{GUE}}
<\sbar{\Lambda}{\mathrm{d}}\).

\begin{figure}[ht]
 \begin{center}
        \includegraphics[width=.8\textwidth]{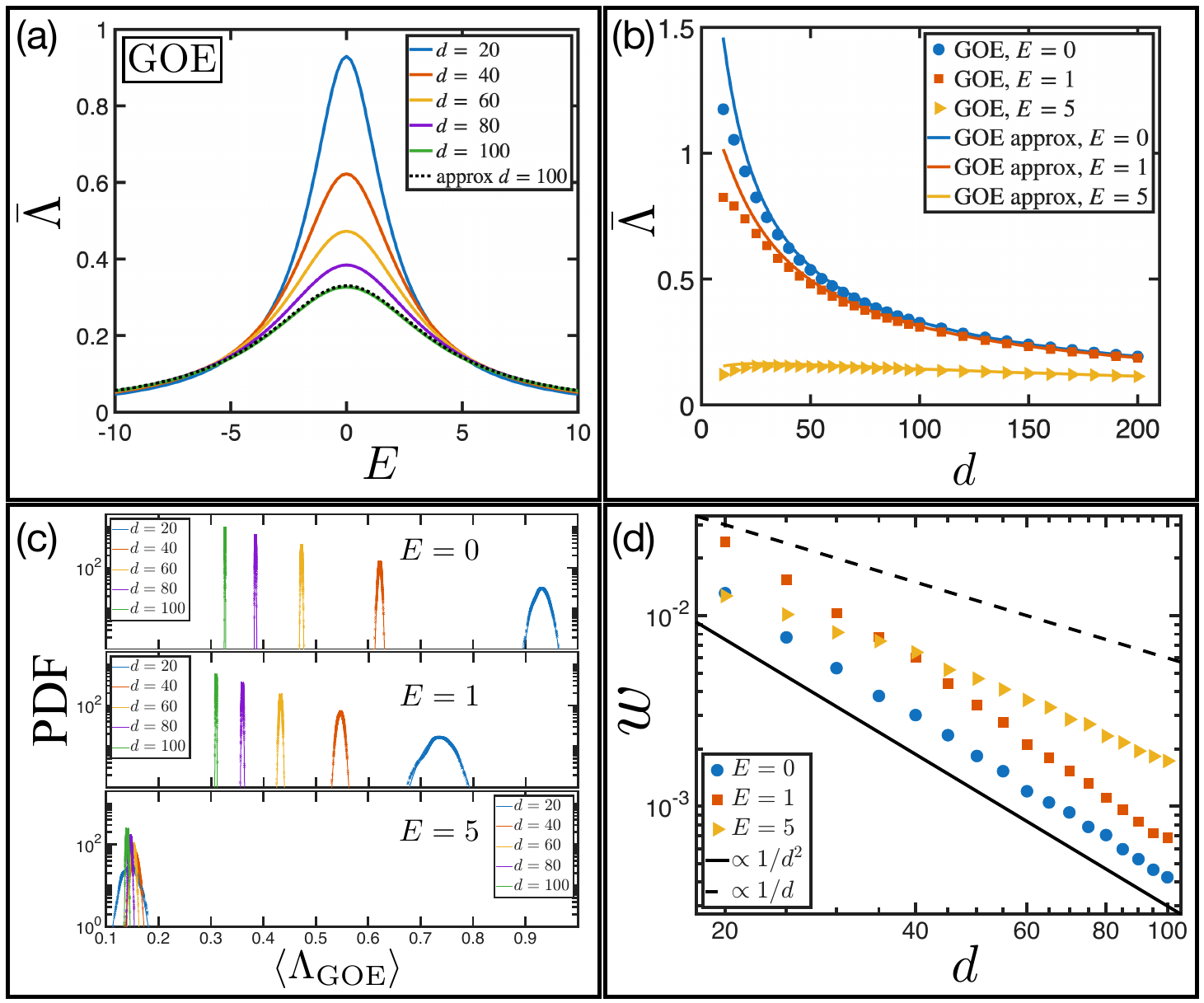}
 \caption{
Numerical simulations for the GOE ensemble.
(a) Sample-averaged mean LE as a function of the energy \(E\) for different
matrix dimensions \(d\) (solid lines).
The sample size varies from 200 realizations for \(d=20\) to 3 realizations
for \(d=100\).
The black dashed line represents the approximation,
\cref{Eq:LambdaGOEGUE}, for \(d=100\).
(b) Sample-averaged mean LE as a function of the matrix dimension \(d\), together with the approximation,
\cref{Eq:LambdaGOEGUE}, for \(E=0\), \(1\), and \(5\).
(c) Probability density function of the sample-averaged mean LE
(logarithmic vertical axis) for different values of \(d\) and \(E\),
obtained from 1000 realizations for each parameter set.
The solid lines are normal-distribution fits,
\(\mathcal{N}(\mu,w^2)\).
(d) Width parameter \(w\) of the normal-distribution fit as a function of
the matrix dimension \(d\) (log-log scale), obtained from the same data set.
The solid and dashed black lines indicate the asymptotic scalings
\(w\propto d^{-2}\) and \(w\propto d^{-1}\), respectively.
}
\label{Fig:GOE_LE}
\end{center}
\end{figure}

In \cref{Fig:GOE_LE}(a), we present the ensemble-averaged LE as a function of
the energy parameter \(E\) for the GOE ensembles of different sizes
(\(d=20,40,60,80,100\)); the black dashed line corresponds to
\cref{Eq:LambdaGOEGUE}.
The dependence on matrix size and a comparison with the approximation are shown
in \cref{Fig:GOE_LE}(b).

The LEs are self-averaging quantities; that is, they concentrate around their
mean as \(V\) increases. This phenomenon has two reasons.
The first is the aforementioned convergence of \(\Gamma_v\) to its mean value
\(\Gamma\).
The second arises from the summation over the elements of \(\Matr{B}\).
Each vertex-scattering block contains \(d(d+1)/2\) distinct matrix elements,
which scales as \(d^2/2\) for \(d\gg1\).
Although these elements are correlated, their correlations weaken as \(1/d\).
One therefore expects the LE to converge to a normal distribution with the mean
given by \cref{Eq:LambdaGOEGUE} and a width \(w\) scaling as
\(1/\sqrt{D}\propto1/d\).
Thus, the width decreases with increasing matrix size.
In practice, for sufficiently large \(d\), a single realization of the
Hamiltonian is already representative of the ensemble, so that
\(\langle\Lambda_{\mathrm{GOE}}\rangle
\to\sbar{\Lambda}{\mathrm{GOE}}\), and similarly for the GUE; see
\cref{Fig:GOE_LE}(c),(d).

It is important to note that the leading term in \cref{Eq:LambdaGOEGUE} can also
be derived using a mean-field approximation, in which the relevant random
quantities involving \(h_{vw}\) and \(H_{vv}\) are replaced by their ensemble
averages.
In the following sections, we will use the mean-field approximation without the
rigorous justification provided here.

\subsubsection{G\(\beta\)E ensembles}\label{Sec:MLE-GbetaE}
The ensembles considered here consist of real symmetric tridiagonal
\(V\times V\) matrices depending on a parameter \(\beta>0\)
\cite{Dumitriu2002}.
For a given \(\beta\), the diagonal elements are
\(H_{nn}=a_n/\sqrt{4\beta V}\), where the \(a_n\) are normally distributed
with zero mean and variance \(2\).
The off-diagonal elements are
\(H_{n-1,n}=H_{n,n-1}=b_n/\sqrt{4\beta V}\), \(n=2,\dots,V\), where the
\(b_n\) are \(\chi_{n\beta}\)-distributed random variables:
\begin{equation}\label{Eq:ChiDistr}
p(b_n)
=
\frac{b_n^{\beta n-1}}
{2^{\frac{\beta n}{2}-1}\Gamma\left(\frac{\beta n}{2}\right)}
e^{-\frac{b_n^2}{2}},
\qquad
b_n\geqslant0.
\end{equation}
The spectral statistics of the G\(\beta\)E ensembles with \(\beta=1,2\)
are identical to those of the GOE and GUE, respectively
\cite{Dumitriu2002}.

The corresponding graph is a chain of \(V\) vertices with connectivity \(d=2\),
except for the two end vertices, which have degree \(1\).
The number of directed edges is \(D=2(V-1)\).
These graphs are bipartite, and the spectrum of their \(\Matr B\) matrix contains
the eigenvalue \(-1\).
The corresponding dynamics is irreducible, since every directed edge can be
reached from every other directed edge, but it is periodic with period \(2\):
the trajectory alternates between the two sublattices and can return to its
initial sublattice only after an even number of steps.
Consequently, the dynamics is not mixing.
To overcome this difficulty, we use the bipartite structure to represent
both \(\Matr U\) and \(\Matr B\) as four square blocks of dimension \(D/2\),
with zero diagonal blocks.
The off-diagonal blocks of \(\Matr U\), denoted by
\(\Matr{U}_{\mathrm{even}}\) and \(\Matr{U}_{\mathrm{odd}}\), contain the
scattering matrices associated with the even and odd vertices, respectively.
The corresponding off-diagonal blocks of \(\Matr B\) are denoted by
\(\Matr{B}_{\mathrm{even}}\) and \(\Matr{B}_{\mathrm{odd}}\).

This representation is obtained by arranging the diagonal blocks of
\(\Matr{\Sigma}(E)\) in \cref{Eq:UPSigma} as
\(\Matr{\sigma}^{(2)},\Matr{\sigma}^{(4)},\dots\), followed by
\(\Matr{\sigma}^{(1)},\Matr{\sigma}^{(3)},\dots\), and choosing the permutation
matrix \(\Matr P\) to exchange each directed edge with its reverse.

The reduced unitary and bistochastic matrices are then defined as
\(\Matr{U}_{\mathrm{red}}
=\Matr{U}_{\mathrm{odd}}\Matr{U}_{\mathrm{even}}\) and
\(\Matr{B}_{\mathrm{red}}
=\Matr{B}_{\mathrm{odd}}\Matr{B}_{\mathrm{even}}\), respectively.
They describe the evolution over two consecutive time steps, thereby eliminating
the even--odd alternation.
The spectra of \(\Matr B\) and \(\Matr{B}_{\mathrm{red}}\) are related by
\begin{equation}\label{Eq:BvsBred}
\det\left[\mu\Matr{I}_{D}-\Matr{B}\right]
=
\det\left[\nu\Matr{I}_{D/2}-\Matr{B}_{\mathrm{red}}\right],
\quad
\nu=\mu^2.
\end{equation}

\begin{figure}[ht]
 \begin{center}
        \includegraphics[width=.9\textwidth]{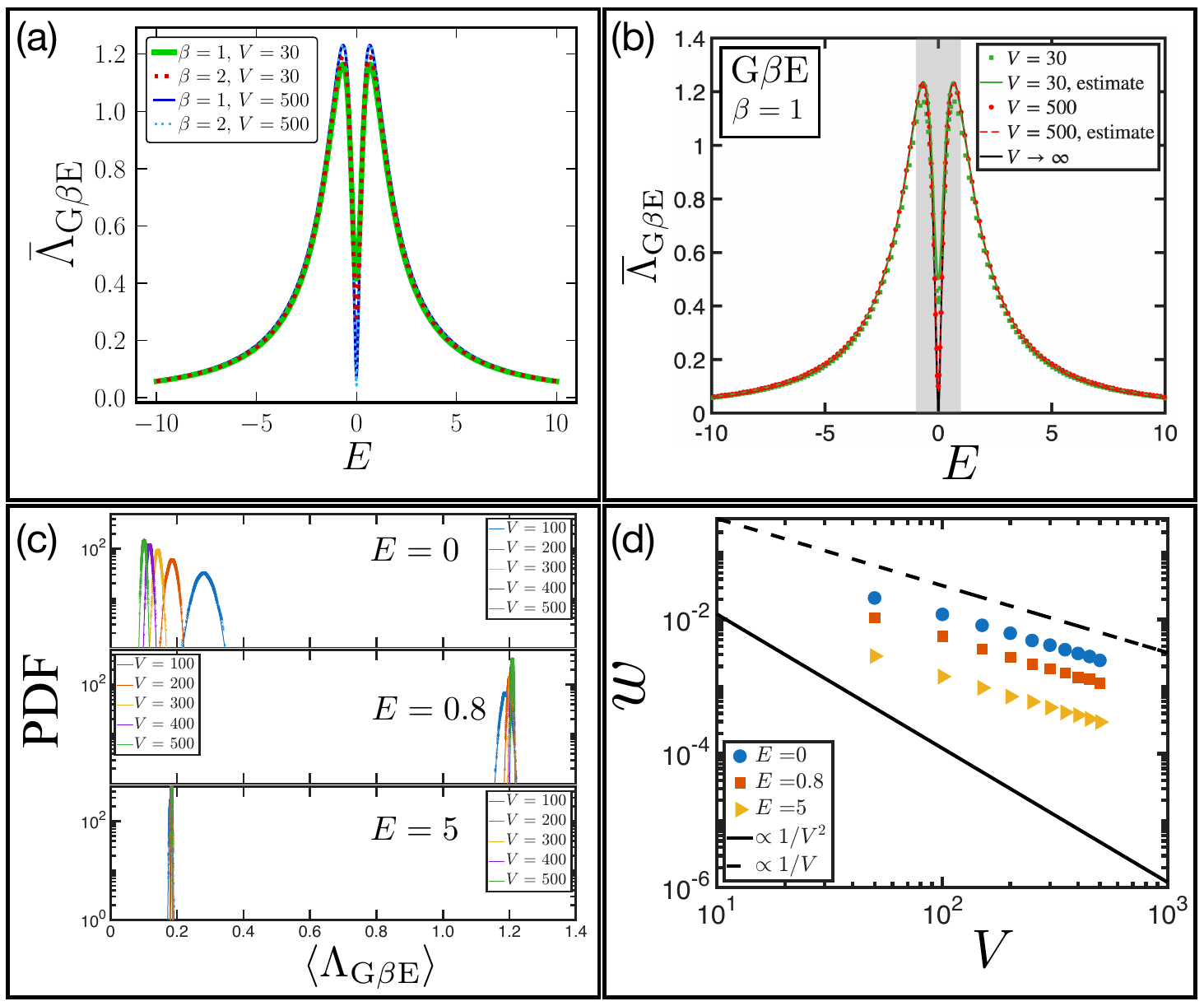}
\caption{
Mean LE of the G\(\beta\)E ensembles.
(a) Sample-averaged mean LE as a function of the energy \(E\) for
\(\beta=1,2\) and \(V=30,500\).
(b) Sample-averaged mean LE as a function of the energy \(E\) for
\(\beta=1\), \(V=30\) (green markers) and \(V=500\) (red markers),
together with the approximation given by \cref{Eq:LE-analyticGbE}.
The black solid line corresponds to the asymptotic limit \(V\to\infty\),
while the shaded region indicates the transition regime.
(c) Probability density function of the sample-averaged mean LE
(logarithmic vertical axis) for different values of \(V\) and \(E\),
obtained from 1000 realizations for each parameter set.
The solid lines are normal-distribution fits,
\(\mathcal{N}(\mu,w^2)\).
(d) Width parameter \(w\) of the normal-distribution fit as a function of the
vertex-space dimension \(V\) (log-log scale), obtained from the same data set.
The solid and dashed black lines indicate the asymptotic scalings
\(w\propto V^{-2}\) and \(w\propto V^{-1}\), respectively.
}
\label{Fig:GbE_LE}
\end{center}
\end{figure}

Numerical simulations, shown in \cref{Fig:GbE_LE}, demonstrate the dependence
of the mean LE and its distribution on \(V\) and \(\beta=1,2\).
The numerical data reveal three main features:
(i) the dependence of the mean LE on \(E\) is similar to that for \(d=2\)
regular graphs;
(ii) for large \(V\), there is hardly any dependence on \(\beta\);
and (iii) there is strong numerical evidence that the mean LE is self-averaging
in the large-\(V\) limit, as in the Wigner--Dyson ensembles.

To reproduce these observations theoretically, we cannot follow the same
approach as for the Wigner--Dyson ensembles for two reasons.
First, the distributions of the matrix elements depend on the position of the
vertex in the graph.
Second, \(\Gamma_v\) is not self-averaging because it is a sum of only two
matrix elements.
We therefore adopt a different theoretical approach, assuming without proof
that the mean-field approximation is valid in the limit of large \(V\).

To estimate the ensemble-averaged LE,
\(\sbar{\Lambda}{\mathrm{G}\beta\mathrm{E}}(E)\), for \(V\gg1\), we use
the mean-field approximation and replace the transmission and reflection
probabilities associated with the \(2\times2\) matrices
\(\Matr{\sigma}^{(n)}\) by their mean values:
\begin{equation}\label{Eq:tnrn}
\sbar{p}{t}(n,E)
\approx
\frac{\frac{n}{V}}
{\frac{n}{V}+E^2+\frac{3}{4\beta V}},
\quad
\sbar{p}{r}(n,E)
\approx
\frac{E^2+\frac{3}{4\beta V}}
{\frac{n}{V}+E^2+\frac{3}{4\beta V}}.
\end{equation}
Introducing the continuous variable \(x=n/V\), we rewrite
\cref{Eq:MeanLE-Th} in an integral form derived in
\cref{SecAppend:GbE-LE}, obtaining the closed-form mean-field expression
\begin{equation}\label{Eq:LE-analyticGbE}
\sbar{\Lambda}{\mathrm{G}\beta\mathrm{E}}
=
2\Big\{
\widetilde{E}^2\Li_{2}(-1/\widetilde{E}^2)
+(\widetilde{E}^2+1)\ln(\widetilde{E}^2+1)
-\widetilde{E}^2\ln\widetilde{E}^2
\left[
\ln(\widetilde{E}^2+1)-\ln\widetilde{E}^2+1
\right]
\Big\},
\end{equation}
where \(\widetilde{E}^2=E^2+3/(4\beta V)\), and
\(\Li_2(\cdot)\) denotes the dilogarithm function.
The leading asymptotic term is independent of \(\beta\). As shown in \cref{Fig:GbE_LE}(b), the prediction of
\cref{Eq:LE-analyticGbE} agrees well with the numerical results.

\begin{figure}[ht]
 \begin{center}
        \includegraphics[width=.6\textwidth]{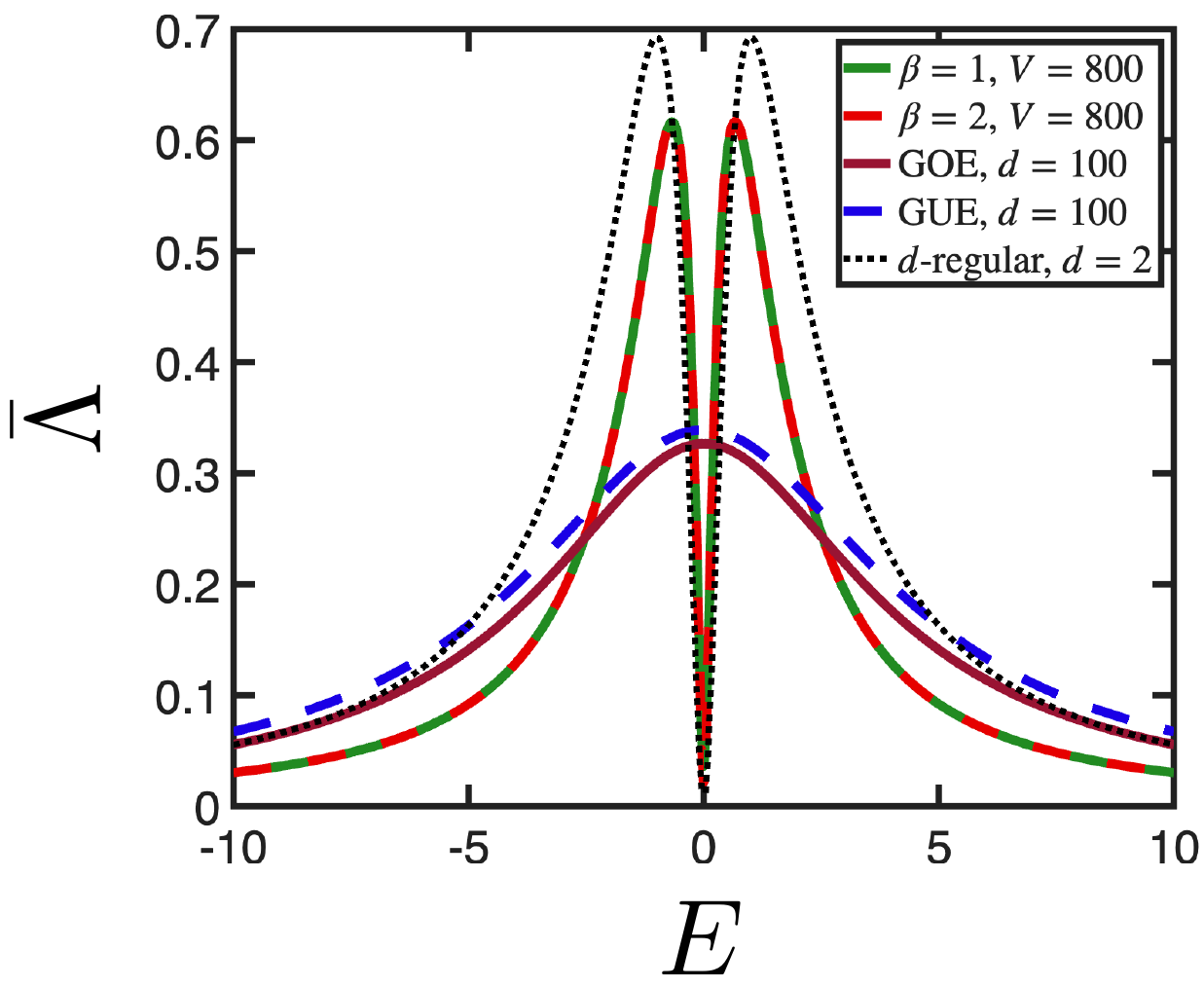}
 \caption{
Comparison of the mean LE for the G\(\beta\)E and Wigner--Dyson ensembles
with that of the \(d\)-regular graph (\(d=2\)).
For graphical clarity,
\(\sbar{\Lambda}{\mathrm{G\beta E}}\) is multiplied by a factor of \(1/2\).
}
\label{Fig:Lambda-vs-E-compare}
\end{center}
\end{figure}

\Cref{Fig:Lambda-vs-E-compare} compares the numerical results for the
Wigner--Dyson and G\(\beta\)E ensembles with those for a \(d\)-regular graph
with \(d=2\).
Although the spectral statistics of the G\(\beta\)E ensembles with
\(\beta=1,2\) are identical to those of the GOE and GUE, respectively, their
mean LEs differ qualitatively.
Moreover, in agreement with \cref{Eq:LambdaGOEGUE},
\(\sbar{\Lambda}{\mathrm{GUE}}\) is slightly greater than
\(\sbar{\Lambda}{\mathrm{GOE}}\).
By contrast, the results for the G\(\beta\)E ensembles with \(\beta=1\) and
\(\beta=2\) are indistinguishable for \(V\gg1\), except in the regime
\(E^2\ll1/V\), in agreement with \cref{Eq:LE-analyticGbE}.
For visual clarity, \(\sbar{\Lambda}{\mathrm{G}\beta\mathrm{E}}\) is
multiplied by a factor of \(1/2\) in the figure.

\subsection{LE variance}\label{Sec:Variance}

Following \cref{Sec:MeanLE}, we discuss the LE variance for each
random-matrix model in the same order.
We begin by evaluating the LE variance defined in
\cref{Eq:LambdaMoments}, which we reproduce here for convenience:
\begin{equation*}
\begin{gathered}
\langle\left[\delta\Lambda(E)\right]^2\rangle
=
\langle\left[\delta\Lambda(E)\right]^2\rangle_F
+
\langle\left[\delta\Lambda(E)\right]^2\rangle_G
-
\langle\Lambda\rangle^2,\\
\langle(\delta\Lambda)^2\rangle_F
=
\frac{1}{D}\sum\limits_{e,e'}B_{e'e}\left[\ln B_{e'e}\right]^2,
\quad
\left\langle(\delta\Lambda)^2\right\rangle_G
=
2\sum\limits_{k\neq0}
\frac{(0|\Matr{G}|r_k)(l_k|\Matr{G}|0)}
{(l_k|r_k)\left(1-\nu_k\right)}.
\end{gathered}
\end{equation*}
As mentioned above,
\(\langle(\delta\Lambda)^2\rangle\) is itself a random variable because it is
associated with a single realization of a random matrix.
Its ensemble average is denoted by
\(\sbar{(\delta\Lambda)^2}{}\), see \cref{Eq:averageVSmean}.

To evaluate the \(G\)-component of the variance and to characterize the
relaxation properties of the Markov dynamics, we analyse the spectrum of
\(\Matr{B}\), including both its eigenvalues and eigenvectors.
In particular, we evaluate the spectral gap
\(1-|\nu_1|\), where \(\nu_1\) is the eigenvalue of the second-largest modulus.
\subsubsection{\(d\)-regular graph}\label{Sec:VarLEdreg}

We start by proving that for this ensemble, the \(G\)-component in
\cref{Eq:LambdaMoments} vanishes identically.
Recall that the matrix \(\Matr{B}\) can be expressed as in \cref{Eq:BPX}, with
the diagonal elements of \(\Matr{\Pi}\) equal to \(p_r\) and all other nonzero
elements equal to \(p_t\), as defined in \cref{Eq:Bd}.
Similarly, the matrix \(\Matr{G}\) can be written as
\(\Matr{G}=\Matr{P}\Matr{Y}\), where \(\Matr{Y}\) has the same structure as
\(\Matr{\Pi}\), with elements \(p_r\ln p_r\) and \(p_t\ln p_t\), respectively,
and \(\Matr{P}\) is the permutation matrix defined in
\cref{Sec:PrelimQuantum}.
We can then write
\[
\Matr{Y} = \Matr{\Pi}\ln\,p_{t} + \Matr{I}p_{r}\ln\left(\frac{p_{r}}{p_{t}}\right), \quad
\Matr{G} = \Matr{B}\ln p_{t} + \Matr{P}\Matr{I}p_{r}\ln\left(\frac{p_{r}}{p_{t}}\right).
\]
If \(p_r=p_t\), the matrices \(\Matr{G}\) and \(\Matr{B}\) commute and share
the same eigenvector space.
In general, however, this need not be the case.
Nevertheless,
\begin{equation}\label{Eq:GzeroProjection}
(0|\Matr{G}|r_{k}) = (0|\Matr{B}|r_{k})\ln p_{t} + (0|\Matr{P}\Matr{I}|r_{k})p_{r}\ln\left(\frac{p_{r}}{p_{t}}\right) = \left[\ln p_{t} + p_{r}\ln\left(\frac{p_{r}}{p_{t}}\right)\right](0|r_{k}) = 0, \quad k\neq 0,
\end{equation}
because \((0|\Matr{P}=(0|\).
Thus, all terms in the sum defining the \(G\)-component vanish.

The \(F\)-component of the variance for \(d\)-regular graphs can be computed
exactly, analogously to \cref{Eq:LambdaDreg}, yielding
\begin{equation}\label{Eq:VarDreg}
\langle\left(\delta\Lambda_{\mathrm{d}}\right)^{2}\rangle = \frac{4(d-1)}{d^{2} + \lambda^{2}}\left(\ln\left[\frac{d^{2} + \lambda^{2}}{4}\right]\right)^{2} +
\frac{(d-2)^{2} + \lambda^{2}}{d^{2} + \lambda^{2}}\left(\ln\left[\frac{d^{2} + \lambda^{2}}{(d-2)^{2} + \lambda^{2}}\right]\right)^{2} - \langle\Lambda_{\mathrm{d}}\rangle^{2},
\end{equation}
where \(\lambda=E/\xi\), and \(\langle\Lambda_{\mathrm{d}}\rangle\) is given by
\cref{Eq:LambdaDreg}.
For large \(d\), the relative standard deviation
\(\sqrt{\langle\left(\delta\Lambda_{\mathrm{d}}\right)^{2}\rangle}/
\langle\Lambda\rangle\)
scales as \(\sqrt{d}\).
As for \(\langle\Lambda_{\mathrm d}\rangle\), the value of
\(\langle(\delta\Lambda_{\mathrm d})^2\rangle\) does not depend on \(V\) or on
the particular realization of the adjacency matrix \(\Matr A^{(\rm d)}\).
Therefore,
\[
\sbar{(\delta\Lambda_{\mathrm d})^2}{}=
\langle(\delta\Lambda_{\mathrm d})^2\rangle.
\]

\begin{figure}[ht]
 \begin{center}
        \includegraphics[width=.9\textwidth]{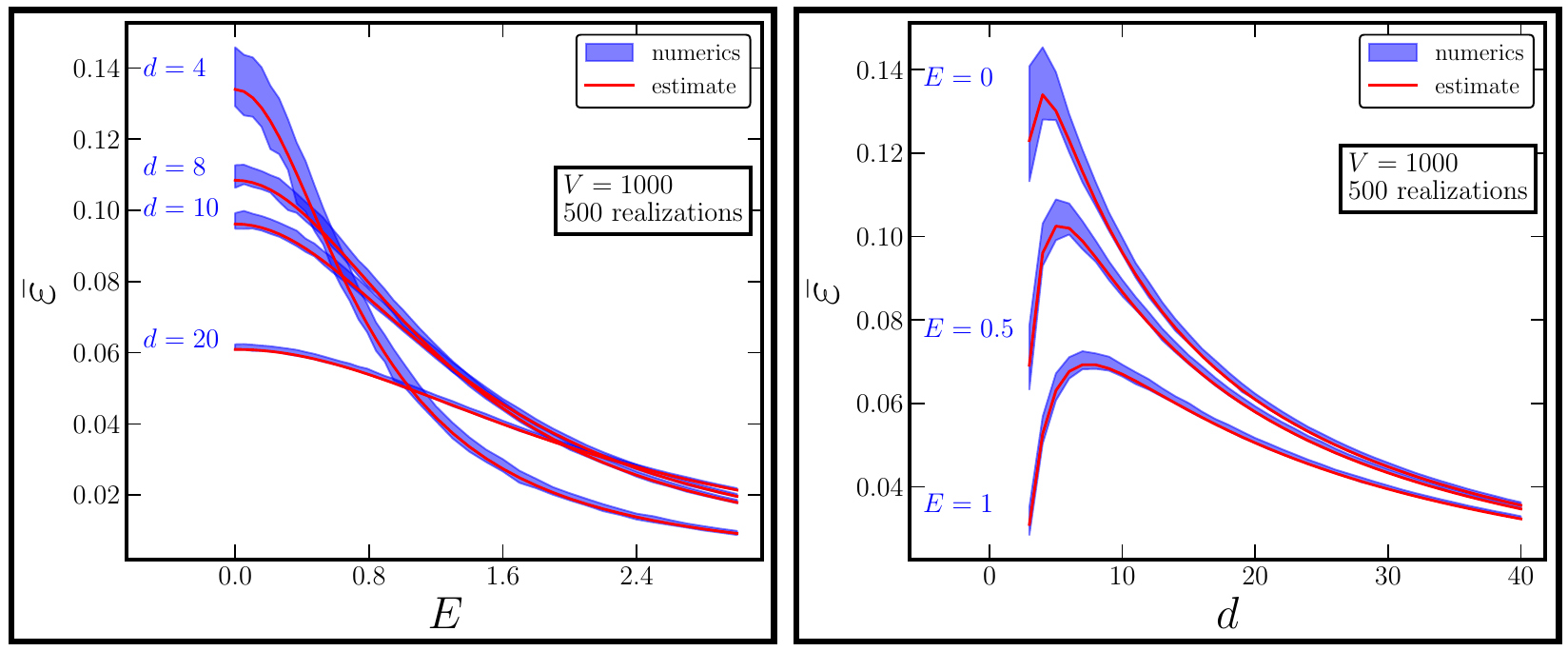}
 \caption{
Spectral gap of the Markov--Poincar\'e map \(\Matr{B}\) associated with a \(d\)-regular
graph, see \cref{Eq:DregGap}.
{\it Left panel:} Ensemble-averaged spectral gap as a function of the energy
\(E\) for different values of \(d\).
{\it Right panel:} Ensemble-averaged spectral gap as a function of \(d\) for
different values of \(E\).
The markers show the numerical simulations, while the lines represent the
lower-bound estimate given by \cref{Eq:DregGap}.
}
\label{Fig:Dreg_gap}
\end{center}
\end{figure}

The spectral gap of \(\Matr{B}\) for a \(d\)-regular graph is a random variable
determined by the nontrivial eigenvalue with the largest modulus of the adjacency
matrix \(\Matr{A}^{\mathrm{(d)}}\).
A lower bound for the spectral gap is
\begin{equation}\label{Eq:DregGap}
\sbar{\varepsilon}{\mathrm{d}}\gtrsim 1 - \frac{2\alpha_{1}}{\lambda^{2} + d^{2}} - \sqrt{\left(\frac{2\alpha_{1}}{\lambda^{2} + d^{2}}\right)^{2} + 1 - \frac{4d}{\lambda^{2} + d^{2}}},\quad \lambda = E/\xi,
\end{equation}
where \(\alpha_1=2\sqrt{d-1}\) is the edge of the asymptotic Kesten--McKay
distribution for the nontrivial eigenvalues of
\(\Matr{A}^{\mathrm{(d)}}\) \cite{McKay1981}.
This expression is illustrated in \cref{Fig:Dreg_gap}.

The spectrum of \(\Matr{B}\) and its relation to the spectrum of
\(\Matr{A}^{\mathrm{(d)}}\), together with the spectral-gap distribution and the
effects of finite \(V\) and non-Ramanujan eigenvalues, will be discussed in a
forthcoming publication \cite{KurnosovSmilansky2026}.
In the present work, we restrict ourselves to this lower bound.

\subsubsection{Wigner-Dyson ensembles}\label{Sec:VarGOE}
\begin{figure}[!htbp]
 \begin{center}
        \includegraphics[width=.8\textwidth]{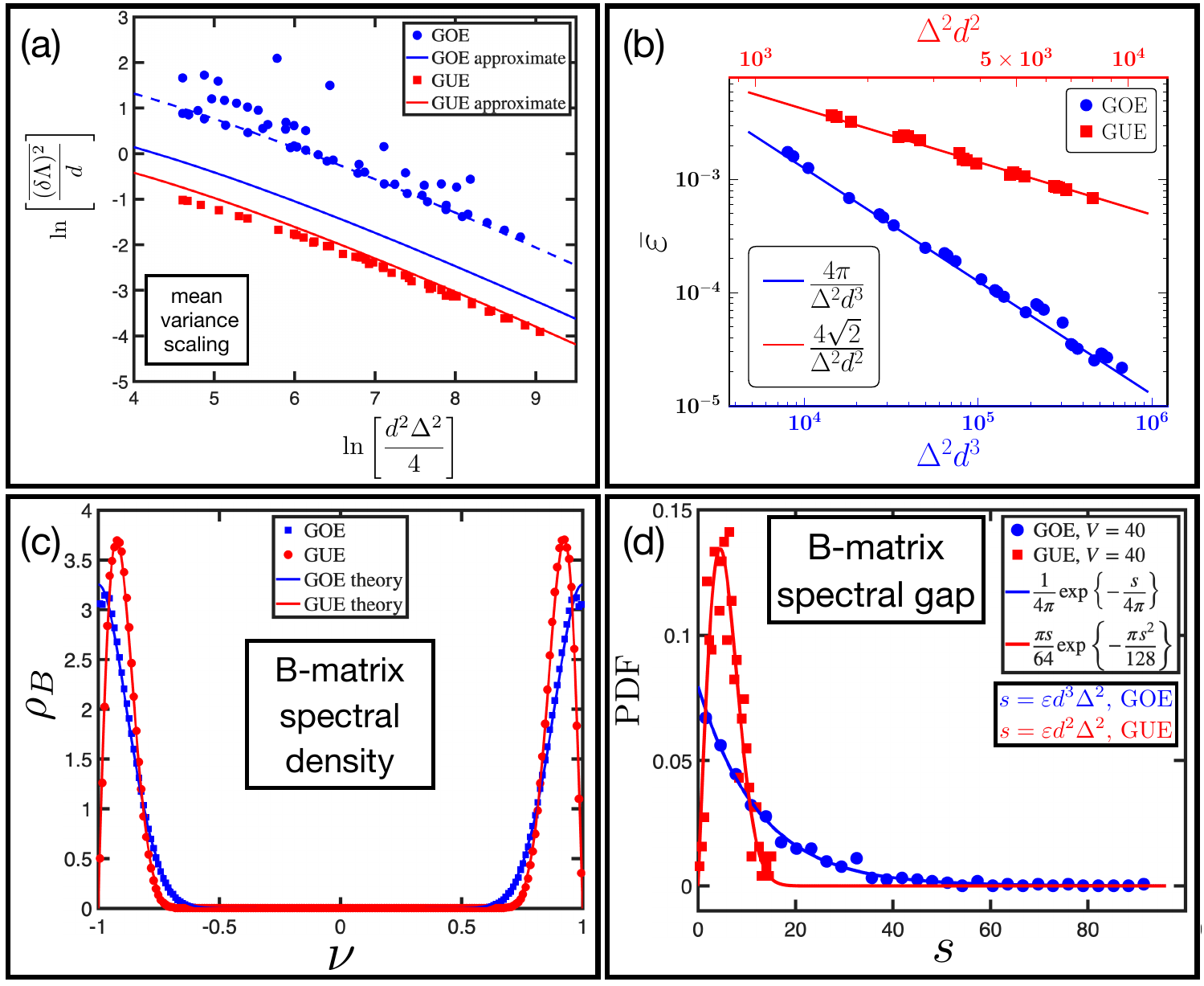}
\caption{
(a) Ensemble-averaged LE variance for the GOE and GUE ensembles, illustrating
the predicted scaling, see \cref{Eq:VarianceWDfinal}.
For the GOE, the blue dashed line is obtained by shifting the theoretical curve
for illustrative purposes.
(b) Sample-averaged spectral gap for the GOE and GUE ensembles, together with
the theoretical predictions, see
\cref{Eq:GapWD}.
(c) Spectral density of the Markov--Poincar\'e map \(\Matr{B}\) for the GOE
and GUE ensembles (\(d=40\)), see \cref{Eq:rho-nu}.
(d) Probability density function of the spectral gap, together with the
theoretical predictions, see \cref{Eq:PDFgapFinal}.
}
\label{Fig:WD_Var_Gap_Distr}
\end{center}
\end{figure}
In \cref{Fig:WD_Var_Gap_Distr}(a) we present the numerical results for
\(\overline{(\delta\Lambda)^2}\) for the GOE (blue circles) and GUE (red
squares) ensembles.
One can see that all the data collapse onto a single curve, although the GOE
exhibits considerably stronger fluctuations.
Figure~\ref{Fig:WD_Var_Gap_Distr}(b) shows the corresponding spectral gap.
Rather than plotting the sample averages as functions of the energy and the
matrix size separately, we introduce appropriate scaling parameters, whose
relevance will be justified below.

First, we consider the \(F\)-component of the LE variance.
It can be estimated following the same procedure used to derive
\cref{Eq:LambdaGOEGUE}.
However, the leading asymptotics can already be obtained within the
mean-field approximation.
Analogously to \cref{Eq:VarDreg}, it reads
\begin{equation}\label{Eq:Fcomponent}
\sbar{(\delta\Lambda)^{2}}{F} =
d\frac{4}{\Delta^{2}d^{2}}
\left[\ln\left(\frac{d^{2}\Delta^{2}}{4}\right)\right]^{2}
+ o\left([\ln d]^{2}/d\right),
\quad
\Delta^{2} = 1 + \left(\frac{E}{\xi}\right)^{2},
\end{equation}
where \(\xi\) is the mean absolute value of the off-diagonal Hamiltonian
matrix element introduced in \cref{Sec:MLE-GOEGUE}.
The centering term
\(\langle\Lambda\rangle^{2}
=
\mathcal{O}\left([\ln d]^{2}/d^{2}\right)\)
can therefore be neglected in the limit \(d\gg1\).

The estimate of the \(G\)-component is more involved, as it depends on both the
spectral density and the eigenspace of \(\Matr{B}\).
As will be shown in \cite{KurnosovSmilansky2026}, the spectral density can be
estimated by
\begin{equation}\label{Eq:rho-nu}
\rho_{B}(\nu) =
\left\{
\begin{aligned}
&\frac{d\Delta^{2}}{4\pi}\exp\left\{-\frac{d^{2}\Delta^{4}(1-|\nu|)^{2}}{16\pi}\right\},
&\text{for GOE},\\
&\frac{\pi d^{2}\Delta^{4}}{64}(1 - |\nu|)\exp\left\{-\frac{\pi d^{2}\Delta^{4}(1 - |\nu|)^{2}}{64}\right\},
&\text{for GUE}.
\end{aligned}
\right.
\end{equation}
Equation~\eqref{Eq:rho-nu} implies that, for \(d\gg1\), the spectrum of
\(\Matr{B}\) is real for both models.
The spectral density near the spectral edges is dramatically
different, as illustrated in \cref{Fig:WD_Var_Gap_Distr}(c).
This difference is neither intuitively obvious nor completely surprising.
Similar differences between GOE and GUE appear in the level-spacing statistics.
Nevertheless,  the Wigner surmise is not directly related to the spectral properties
of the matrices \(\Matr{B}\).

Another difference between the GOE and GUE spectra is the probability density
function of the spectral gap,
\(\varepsilon = 1 - |\nu_{1}|\), shown in
\cref{Fig:WD_Var_Gap_Distr}(d).
It is given by
\begin{equation}\label{Eq:PDFgapFinal}
p_{\mathrm{gap}}(s)  =
\left\{
\begin{aligned}
&\frac{1}{4\pi}\exp\left(-\frac{s}{4\pi}\right), &s = \varepsilon d^{3}\Delta^{2},\, &\text{GOE},\\
&\frac{\pi s}{64}\exp\left(-\frac{\pi s^{2}}{128}\right),&s = \varepsilon d^{2}\Delta^{2},\, & \text{GUE},
\end{aligned}
\right.
\end{equation}
The corresponding estimate for the mean spectral gap is
\begin{equation}\label{Eq:GapWD}
\sbar{\varepsilon}{\mathrm{WD}} =
\left\{
\begin{aligned}
&\frac{4\pi}{\Delta^{2}d^{3}},& \text{GOE},\\
&\frac{4\sqrt{2}}{\Delta^{2}d^{2}},& \text{GUE}.
\end{aligned}
\right.
\end{equation}
The GOE ensemble exhibits two distinctive features.
First, the gap PDF remains finite as \(\varepsilon\to0\).
Second, the mean gap scales as
\(\mathcal{O}(d^{-3})\), in contrast to the
\(\mathcal{O}(d^{-2})\) scaling of the GUE.
Equation~\eqref{Eq:GapWD} justifies the choice of scaling parameters in
\cref{Fig:WD_Var_Gap_Distr}(b).

Having established the spectral properties of \(\Matr{B}\), we now evaluate
the contribution of the \(G\)-component to the variance.
The \(G\)-component can be estimated using a slightly modified mean-field
approximation previously employed for the Wigner--Dyson ensembles.
As will be shown in \cite{KurnosovSmilansky2026}, the \(G\)-component exhibits
the same leading-order scaling as the \(F\)-component in
\cref{Eq:Fcomponent}.
Consequently, the full ensemble-averaged LE variance is given by
\begin{equation}\label{Eq:VarianceWDfinal}
\sbar{(\delta\Lambda)^{2}}{} \approx
d\frac{4}{d^{2}\Delta^{2}}
\left[\ln\left(\frac{\Delta^{2}d^{2}}{4}\right)\right]^{2}
(1 + 2 C),
\quad
C =
\left\{
\begin{aligned}
&\frac{\pi^{2} - 4}{4}, \, &\text{GOE},\\
&\frac{16 - \pi^{2}}{\pi^{2}}, \, &\text{GUE}.
\end{aligned}
\right.
\end{equation}

One can see the leading-order scaling of the ensemble-averaged LE variance for
the Wigner--Dyson ensembles in
\cref{Fig:WD_Var_Gap_Distr}(a).
For the GUE, the agreement between theory and numerics can be further improved
by a more careful evaluation of the next-order asymptotics, as was done for the
mean LE in \cref{Eq:LambdaGOEGUE}.
For the GOE, however, \cref{Eq:VarianceWDfinal} correctly reproduces only the
scaling.
Its prefactor provides a lower bound to the numerical data (solid blue line),
whereas the blue dashed line is obtained by a vertical shift of the theoretical
curve in order to highlight the predicted scaling.
In addition, the GOE data exhibit substantially larger fluctuations about the
scaling curve than those of the GUE.

The origin of these unusually large fluctuations can be understood from the
spectral statistics of \(\Matr{B}\).
Indeed, according to \cref{Eq:rho-nu}, the mean inverse spectral gap,
\(\langle(1-\nu)^{-1}\rangle\), diverges for the GOE, although the mean gap
\(\langle1-\nu\rangle\propto d^{-1}\) remains finite.
This indicates that, in the limit \(1-\nu\to0\), the small denominator in
\cref{Eq:LambdaMoments} is compensated by correspondingly small projections of
\(\Matr{G}\) onto the associated eigenvectors of \(\Matr{B}\), ensuring the
convergence of the \(G\)-component.
Nevertheless, an imperfect cancellation between these quantities may produce
exceptionally large contributions to the \(G\)-component, leading to
fluctuations far beyond those expected from a normal distribution.
Such sensitivity to eigenvalues near the spectral edge is a generic consequence
of \cref{Eq:LambdaMoments}. It therefore suggests algebraic tails in the
probability density function of
\(\langle(\delta\Lambda)^2\rangle\).
The GOE is exceptional because the finite spectral density near
\(\nu\simeq1\) makes these rare events significantly more frequent.

\begin{figure}[!htbp]
 \begin{center}
        \includegraphics[width=.7\textwidth]{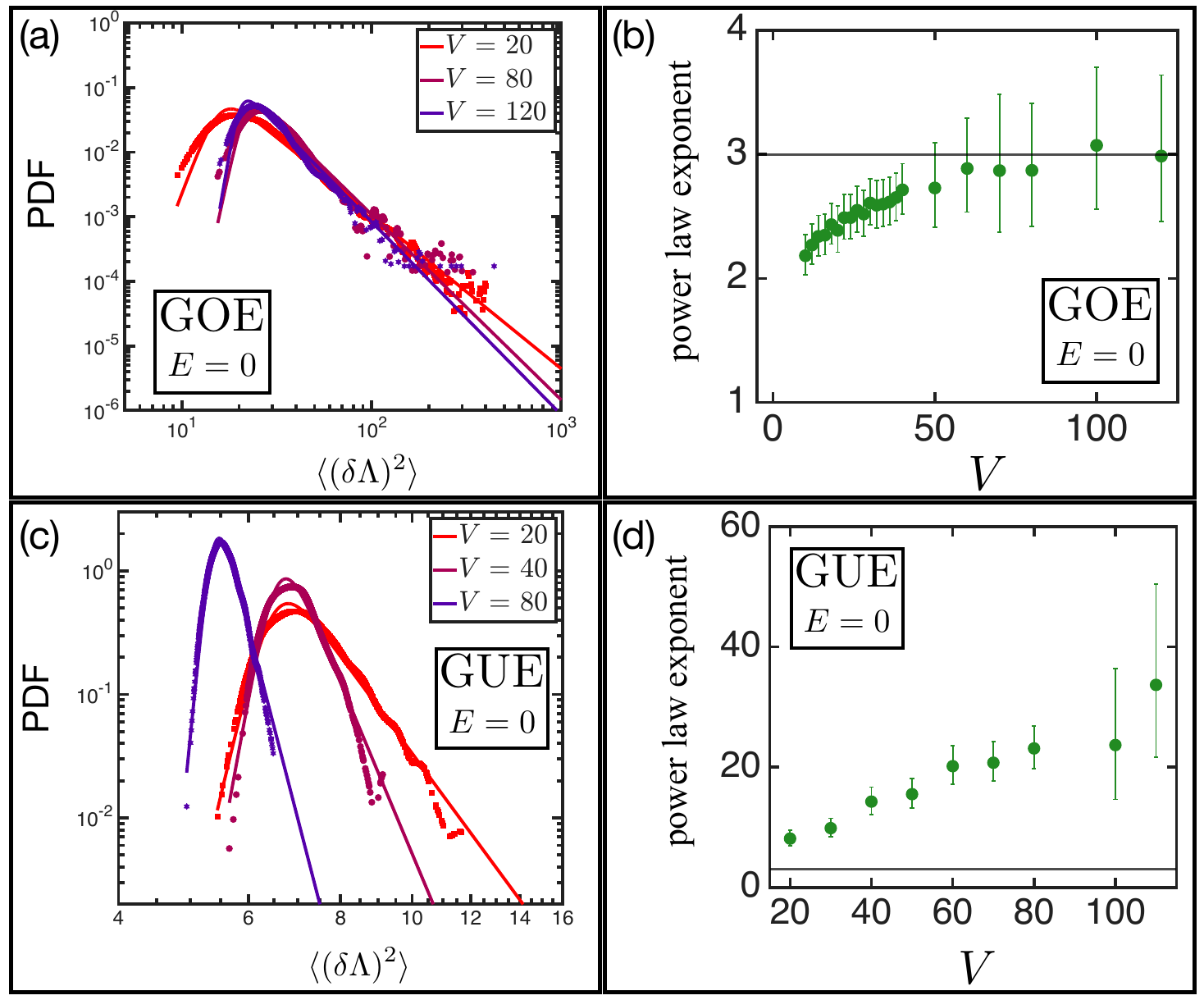}
 \caption{(a) Probability density function of the LE variance,
\(\langle(\delta\Lambda)^2\rangle\), for the GOE ensemble at \(E=0\),
together with the Burr--XII fit, \cref{Eq:Burr}.
(b) Power-law exponent \(ck+1\) extracted from the Burr fit as a function of
the matrix dimension \(V\).
The exponent remains close to the critical value \(3\), indicating a heavy-tailed
distribution.
(c), (d) Same as (a), (b), respectively, for the GUE ensemble.
In contrast to the GOE, the power-law exponent is substantially larger than
\(3\) and increases with \(V\).}
\label{Fig:WD_Burr}
\end{center}
\end{figure}

So far we have focused on the ensemble average of the LE variance.
To obtain a better understanding of its distribution and to assess whether it
is a self-averaging quantity, we performed detailed numerical simulations.
In particular, we computed the probability density function of the LE variance
and investigated its dependence on the energy \(E\) and the matrix dimension
\(V\).
The resulting distributions are shown in
\cref{Fig:WD_Burr}(a),(c).
They share two common features.
First, they are centered around mean values that agree with the mean-field
predictions.
Second, they exhibit algebraic tails.

Since no theoretical expression for the variance PDF is currently available, we
fit the numerical data using the Burr--XII distribution \cite{Burr1942}, which provides an
excellent description of the observed PDFs.
It is a three-parameter distribution with probability density
\begin{equation}\label{Eq:Burr}
p_{\mathrm{burr}}(x; a, c, k) =
\frac{k c}{a}
\frac{\left(x/a\right)^{c-1}}
{\left[1 + \left(x/a\right)^{c}\right]^{k+1}},
\quad
x \geqslant 0.
\end{equation}
For \(c\geqslant1\) the PDF has a maximum at
\[
x_{\mathrm{max}}
=
a\left(\frac{c-1}{ck+1}\right)^{1/c},
\]
while for \(c>2\) it also has a minimum at \(x=0\).

In the limit \(x/a\gg1\),
\cref{Eq:Burr} exhibits a power-law decay,
\(\propto x^{-(ck+1)}\).
Consequently, only moments of degree smaller than \(ck\) exist.
Indeed, the non-central moments are given by
\begin{equation}\label{Eq:BurrMoments}
\langle x^n\rangle
=
\frac{a^n}{\Gamma(k)}
\Gamma\left(k-\frac{n}{c}\right)
\Gamma\left(1+\frac{n}{c}\right),
\quad
n<ck.
\end{equation}
In particular, the second moment exists only if the power-law exponent
\(ck+1\) exceeds \(3\).
For \(2<ck+1<3\), the mean is finite, whereas the sample mean becomes an
unreliable estimator of the ensemble average because the second moment is
undefined.

These numerical results support the qualitative picture developed above.
As one can see in \cref{Fig:WD_Burr}(b), the power-law exponent for the GOE
remains close to the critical value of \(3\).
Consequently, the variance PDF possesses a heavy tail, explaining the limited
applicability of \cref{Eq:VarianceWDfinal}.
The GUE behaves very differently, as shown in
\cref{Fig:WD_Burr}(d).
Its power-law exponent is substantially larger than \(3\) and exhibits a clear
growth with increasing matrix dimension.


\subsubsection{G\(\beta\)E ensembles}\label{Sec:VarGbE}
\begin{figure}[!htbp]
 \begin{center}
        \includegraphics[width=.7\textwidth]{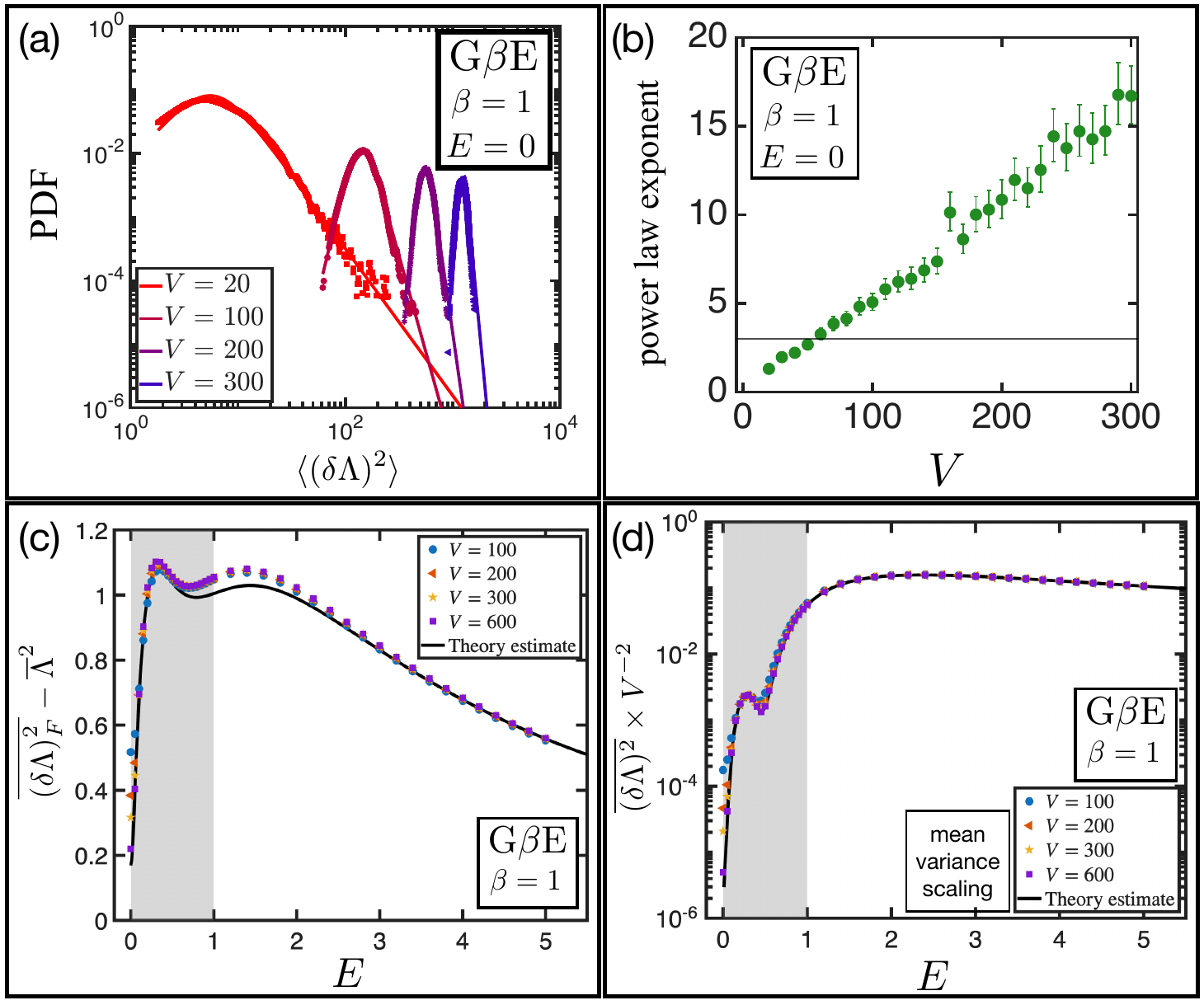}
 \caption{
(a) Probability density function of the LE variance,
\(\langle(\delta\Lambda)^2\rangle\), for the G\(\beta\)E ensemble
(\(\beta=1\), \(E=0\)), together with the Burr--XII fit,
\cref{Eq:Burr}.
(b) Power-law exponent \(ck+1\) extracted from the Burr fit as a function of
the matrix dimension \(V\).
The exponent remains well above the critical value \(3\), indicating that the
sample average is a reliable estimator of the ensemble average.
(c) Ensemble-averaged \(F\)-component of the LE variance together with the
centering term, \(\overline{\Lambda}^{\,2}\), as functions of the energy
\(E\) for different values of \(V\).
(d) Ensemble-averaged LE variance as a function of \(E\) for different values
of \(V\), normalized by \(V^2\).
The black solid line is the theoretical prediction given by the sum of
\(\sbar{(\delta\Lambda)^{2}}{F} - \overline{\Lambda}^{2}\) and \cref{Eq:GcomponentHydroExact}.
}
\label{Fig:GbE_Var}
\end{center}
\end{figure}
The LE variance of the G\(\beta\)E ensembles (independently of the value of
\(\beta\)) also follows the Burr distribution; see, for example,
\cref{Fig:GbE_Var}(a),(b).
Unlike the GOE, however, the power-law exponent is substantially larger than
\(3\), making the sample average a reliable estimator of the ensemble average.

Our numerical simulations for \(V\gg1\) lead to several observations.
First, the ensemble-averaged LE variance is nearly independent of \(\beta\).
Second, the \(F\)-component does not depend on \(V\) and remains of the same
order as the centering term, \(\overline{\Lambda}^{\,2}\).
Third, the \(G\)-component scales as \(V^{2}\).
Consequently, the variance is dominated by the \(F\)-component in the low-energy
regime \(E^{2}V\ll1\), whereas its contribution becomes negligible at larger
energies.
Finally, the LE variance is a non-monotonic function of the energy on the
interval \(E\in[0,1]\).

The numerical results for \(\beta=1\) are presented in
\cref{Fig:GbE_Var}(c),(d), together with estimates obtained within the
mean-field approximation introduced in \cref{Sec:MLE-GbetaE}, although without
rigorous justification.
As one can see, this approximation provides a surprisingly accurate estimate of
the LE variance, just as it did for the mean LE.

\begin{figure}[!htbp]
 \begin{center}
        \includegraphics[width=.6\textwidth]{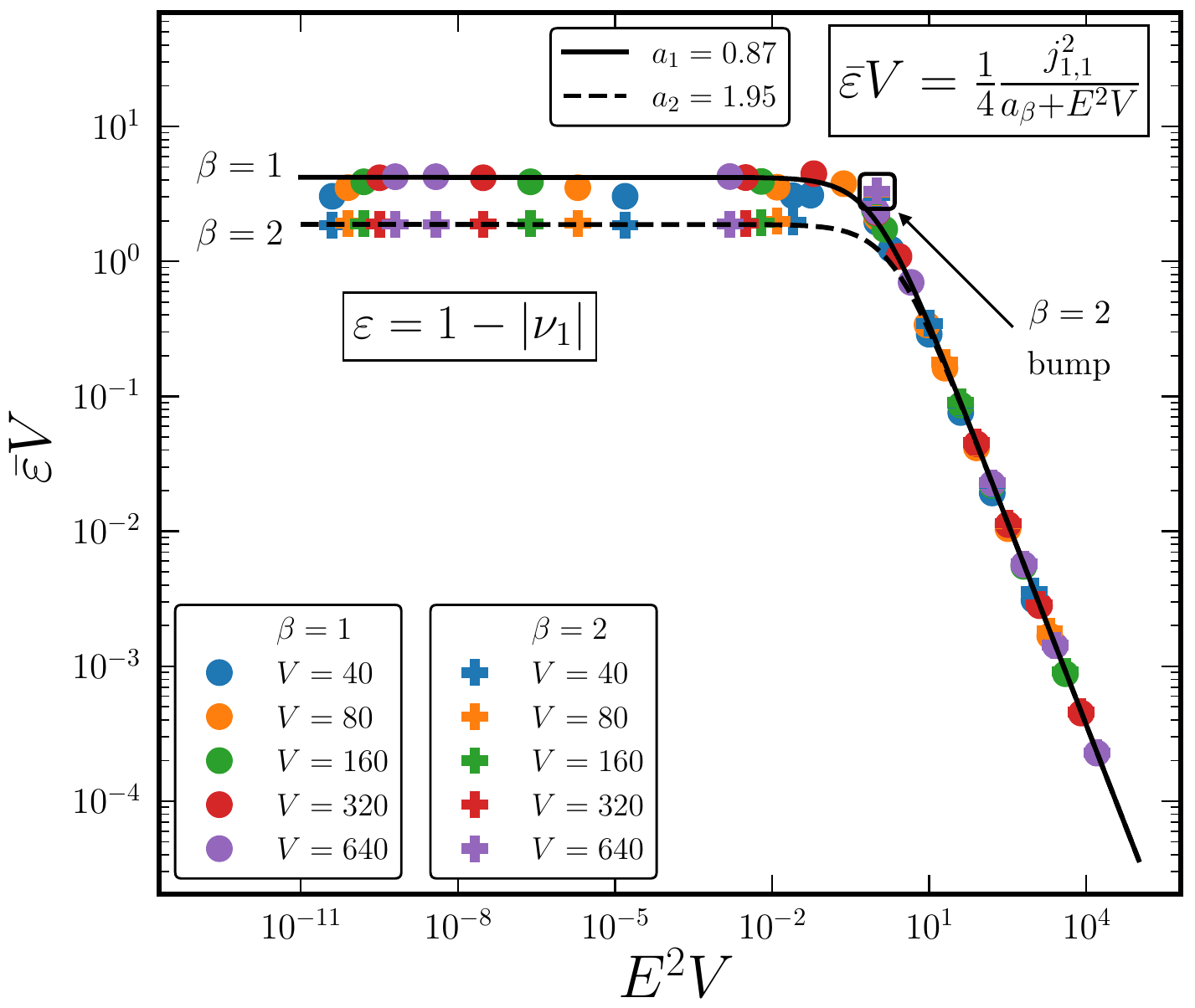}
\caption{
G\(\beta\)E spectral gap scaling.
The symbols show the sample-averaged spectral gap obtained from numerical
simulations, while the lines represent the theoretical prediction for the
ensemble average, \cref{Eq:GbEgapfull}.
}
\label{Fig:GbE-GapScaling}
\end{center}
\end{figure}

In addition to the LE variance, \cref{Fig:GbE-GapScaling} shows the average
spectral gap, \(\overline{\varepsilon}\).
It scales as \(V^{-1}\) for \(E^{2}V<1\) and as
\((EV)^{-2}\) for \(E^{2}V>1\).
The vicinity of \(E^{2}V\sim1\) is a transition region, where the less
fluctuating \(\beta=2\) ensemble exhibits a pronounced bump.
Furthermore, the numerical simulations suggest that this transition region
coincides with the point at which the eigenvalue of largest modulus,
\(\nu_{1}\), becomes real, while the remainder of the spectrum is still
predominantly composed of complex-conjugate pairs.

In the following, we describe the approximate theory that reproduces the
numerical results.

The estimate of the \(F\)-component is straightforward, as it relies entirely
on the approximations introduced in \cref{Sec:MLE-GbetaE}.
The resulting expression, together with its derivation, is rather lengthy and
is therefore presented in \cref{SecAppend:GbE-Fcomponent}.
It is represented by the black solid line in \cref{Fig:GbE_Var}(c).

The theoretical estimates of the \(G\)-component and the spectral gap require
solving the eigenproblem for \(\Matr{B}\).
Here, we quote only the final results of a continuum mean-field approximation;
the transfer-matrix derivation and a detailed analysis of the spectrum will be
presented in a forthcoming publication \cite{KurnosovSmilansky2026}.
In the regime \(E^2V\gg1\), the slow modes are described by an effective
diffusion problem, leading to
\begin{equation}\label{Eq:GcomponentHydroExact}
\sbar{\left(\delta\Lambda\right)^2}{G}
=
V^2
\sum_{k\geqslant1}
\frac{8E^2}{j_{1,k}^2}
\frac{|C_k(E)|^2}
{J_0^2(j_{1,k})},
\qquad
C_k(E)
=
-\frac12
\int\limits_0^1
J_0(j_{1,k}\sqrt{x})
g(x;E)\,dx,
\end{equation}
where \(g(x;E)\) is the continuum counterpart of
\(\Matr{G}|0)\), defined in \cref{Eq:LambdaIntegrand}, \(J_0\) is the Bessel
function of the first kind of order zero, and \(j_{1,k}\) is the \(k\)-th zero
of \(J_1\).
This result explains the \(V^2\) scaling of the \(G\)-component and, together
with the \(F\)-component after subtraction of the centering term, provides the
estimate shown by the black solid line in \cref{Fig:GbE_Var}(d).
Although \cref{Eq:GcomponentHydroExact} is not applicable when \(E^2V\ll1\),
the total LE variance remains accurately estimated because it is dominated by
the \(F\)-component in this regime.

The same continuum analysis motivates the following semi-empirical expression
for the ensemble-averaged spectral gap:
\begin{equation}\label{Eq:GbEgapfull}
\sbar{\varepsilon}{\mathrm{G\beta E}}
=
\frac{1}{4}
\frac{j_{1,1}^{2}}{(a_{\beta}+E^{2}V)V}.
\end{equation}
Here, \(a_\beta\) is determined theoretically from the low-energy solution of
the same continuum eigenproblem, yielding \(a_1\simeq0.87\) and
\(a_2\simeq1.95\).
The full continuum analysis also predicts qualitatively the crossover feature
observed near \(E^2V\sim1\), including the pronounced bump for \(\beta=2\).
The compact interpolation \cref{Eq:GbEgapfull}, however, reproduces only the
two asymptotic scaling regimes and does not include this local feature.
It is represented by the black solid and dashed lines in
\cref{Fig:GbE-GapScaling} for \(\beta=1\) and \(\beta=2\), respectively.
We use \cref{Eq:GbEgapfull} in \cref{Sec:ThermalAvSpectralGap} to calculate the
thermal average of the spectral gap.

\section{Thermal average}\label{Sec:ThermalAverage}

So far, we have considered all observables as functions of the energy.
Here we study their thermal averages as functions of the temperature.
In this section we present the thermal averages of the mean Lyapunov exponent
and the spectral gap of the Markov--Poincar\'e map.

\begin{figure}[!htbp]
 \begin{center}
        \includegraphics[width=.9\textwidth]{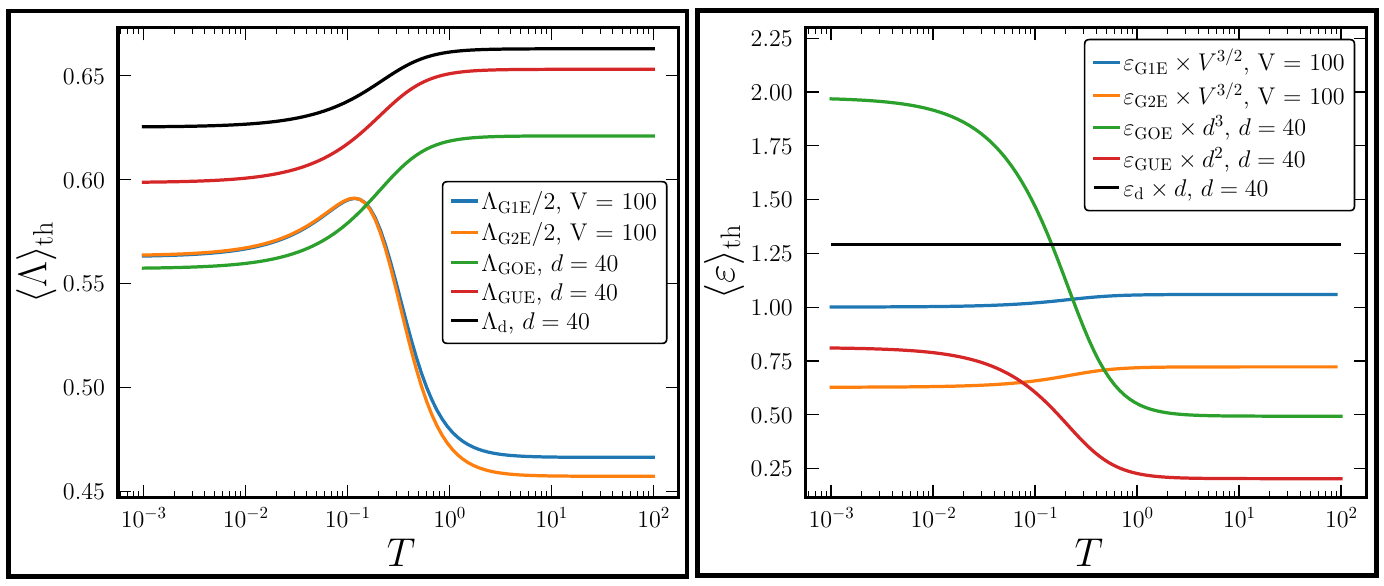}
 \caption{
{\it Left panel}: normalized thermal average of the mean LE,
see \cref{Eq:ThermalAverageGOEGUE,Eq:ThermalAverageGbE,Eq:LEthermalDregGen}.
The thermal average of the G\(\beta\)E ensembles is divided by \(2\) for
graphical clarity.
{\it Right panel}: normalized thermal average of the spectral gap,
see \cref{Eq:ThermalAverageGapWD,Eq:GbE-Gap-Thermal,Eq:DregGap-vs-d}.
The thermal average of the G\(\beta\)E spectral gap is divided by
\(j_{1,1}^{2}/2\).
}
\label{Fig:ThermalAverageEverything}
\end{center}
\end{figure}

The thermal average of an observable \(X\) is defined as
\[
\langle X(T)\rangle_{\mathrm{th}}
=
\frac{1}{\mathcal{Z}(T)}
\int dE\,\rho(E)X(E)e^{-E/T},
\qquad
\mathcal{Z}(T)
=
\int dE\,\rho(E)e^{-E/T},
\]
where \(X=\Lambda\) or \(X =  \varepsilon\), and \(\rho(E)\) is the spectral
density of the corresponding ensemble.
The expressions for \(\overline{\Lambda}(E)\) and
\(\overline{\varepsilon}(E)\) for each ensemble are derived in
\cref{Sec:MeanLE,Sec:Variance}; the spectral density is given by the
Kesten--McKay distribution for \(d\)-regular graphs and by the semicircle
distribution for the Wigner--Dyson and G\(\beta\)E ensembles.
The corresponding integrals are evaluated in
\cref{SecAppend:ThermalAverage}, and the results are presented in
\cref{Fig:ThermalAverageEverything}.

The thermal averages of the LE for all five ensembles are shown in the left
panel of \cref{Fig:ThermalAverageEverything}.
As in the case of \(E=0\), the thermal averages for the Wigner--Dyson
ensembles and the \(d\)-regular graph (for the same values of \(d\)) satisfy
\[
\langle\Lambda_{\mathrm{GOE}}\rangle_{\mathrm{th}}
<
\langle\Lambda_{\mathrm{GUE}}\rangle_{\mathrm{th}}
<
\langle\Lambda_{\mathrm{d}}\rangle_{\mathrm{th}}
\]
throughout the entire temperature range.
The factor of \(1/2\) for
\(\langle\Lambda_{\mathrm{G\beta E}}\rangle_{\mathrm{th}}\) is introduced
primarily for graphical convenience, although it may also be justified by
considering the phase-space reduction, see
\cref{Sec:MLE-GbetaE}.
For the G\(\beta\)E ensembles we choose a moderate value of \(V=100\) in order
to distinguish the cases \(\beta=1\) and \(\beta=2\).
As discussed in \cref{Sec:MLE-GbetaE}, the difference between these two
ensembles vanishes in the large-\(V\) limit; see, for example, the numerical
results in \cref{Fig:GbE_LE}(a).
At present, we cannot offer a physical explanation for the non-monotonic
behaviour of
\(\langle\Lambda_{\mathrm{G\beta E}}\rangle_{\mathrm{th}}\).

The thermal averages of the spectral gap are shown in the right panel of
\cref{Fig:ThermalAverageEverything}.

The thermal average of the LE is asymptotically independent of the system size.
For regular graphs and the Wigner--Dyson ensembles, it depends only weakly on
the connectivity, scaling as \(\ln d/d\)
(for the G\(\beta\)E ensembles the effective connectivity is always \(2\)).
By contrast, the thermal average of the spectral gap exhibits a much stronger
dependence on the system size.
For the G\(\beta\)E ensembles it scales as
\(\mathcal{O}(V^{-3/2})\) and remains sensitive to the value of \(\beta\),
whereas the GOE, GUE, and \(d\)-regular graph each exhibit their own distinct
scaling behaviour.
Overall, for all ensembles considered in the present work,
\[
\langle\varepsilon\rangle_{\mathrm{th}}
\ll
\langle\Lambda\rangle_{\mathrm{th}},
\]
as expected from their respective roles in the system dynamics.

\section{Conclusion}
In the present work, we studied the evolution of quantum systems defined on
arbitrarily large but finite-dimensional vector spaces.
Once a basis is chosen, the Hamiltonian is represented by a matrix.
Following previous work \cite{gnutzmann2024}, we construct a quantum
Poincar\'e map (QPM) for such a system.
The QPM is determined by the original Hamiltonian matrix and the system energy
\(E\), and is represented by a unitary matrix \(\Matr{U}(E)\) acting on an
extension of the original vector space.
The quantum evolution is described by successive applications of
\(\Matr{U}(E)\), with the integer \(t\) counting the number of steps, in analogy
with a Poincar\'e map in classical mechanics.
Trajectories in this extended space are sequences of indices corresponding to
nonvanishing contributions to \(\Matr{U}^t\).

Here, we focused on the semiclassical evolution associated with
\(\Matr{U}(E)\), defined by the bistochastic matrix \(\Matr{B}(E)\), whose
elements are \(B_{ij}(E)=|U_{ij}(E)|^2\).
The matrix \(\Matr{B}(E)\) governs the stochastic evolution of probability
vectors with components \(p_i=|a_i|^2\), where \(a_i\) are the quantum
amplitudes in the space on which \(\Matr{U}(E)\) acts.
Ergodic theory allows us to associate an energy-dependent Lyapunov exponent
with each trajectory and to determine its mean and variance over the full set
of trajectories.
For mixing dynamics, the stochastic evolution approaches the uniform
equilibrium state, with the rate of convergence determined by the spectral gap
between the Perron--Frobenius eigenvalue and the eigenvalue of second-largest
modulus.

We applied the formalism to five random-matrix ensembles: the Wigner--Dyson
Gaussian ensembles GOE and GUE; the Gaussian \(\beta\)-ensembles
(G\(\beta\)E) with \(\beta=1,2\), represented by the Dumitriu--Edelman
tridiagonal matrix model; and the ensemble of adjacency matrices of
\(d\)-regular graphs on \(V\) vertices. The main conclusions can be summarized as follows:
\begin{enumerate}
\item The mean Lyapunov exponent depends on the representation of the Hamiltonian,
rather than on its spectral statistics alone.
\item The mean Lyapunov exponent is self-averaging: as the matrix dimension
increases, its distribution over the ensemble becomes narrower around the
ensemble average.
\item The LE variance is generally not self-averaging and may depend on the
dimension of the matrices in the ensemble.
\item The initial entropy growth rate is the same for the quantum and
semiclassical evolutions and is equal to the mean Lyapunov exponent.
\item For quantum systems consisting of \(N\) spins, each interacting with a
finite number of neighboring spins through random matrix elements, the
ensemble-averaged Lyapunov exponent is independent of the matrix dimension,
which grows exponentially with \(N\), but is proportional to \(N\).
Examples include the random Heisenberg model on a finite linear lattice and
the Parisi hypercube model of spins interacting randomly with their neighbors
on the hypercube.
\end{enumerate}

Several open questions will be addressed in the forthcoming publications:
\begin{enumerate}
\item Quantum corrections to the semiclassical evolution can be studied by
constructing a Poincar\'e map from powers of the quantum evolution operator,
\(B^{(k)}_{ij}=|(\Matr{U}^{k})_{ij}|^{2}\), \(k\geqslant3\), thereby including
interference between orbits of length \(k\geqslant3\).
\item The growth of the Shannon and second R\'enyi entropies under the quantum
evolution of the occupation probabilities, and their relation to the
semiclassical stochastic evolution, remain to be studied in detail.
\item The method presented here is based on the energy-dependent quantum
Poincar\'e map.
To relate topological the time to the physical time, one may either use the
Wigner--Smith method to calculate the delay time per interaction or obtain the
discrete-time evolution by Fourier transforming \cref{Eq:resolvent_identity}.
\item In the context of many-body quantum chaos and quantum information scrambling,
it would be interesting to consider fast scramblers, such as SYK models, and
slow scramblers, such as spin chains.
The relevant dynamical correlation functions and entropies, including OTOCs
and the entanglement entropy of a subsystem, may be represented in terms of the
discrete QPM using the propagator identity \cref{Eq:resolvent_identity}.
One may then ask how classical properties of the map \(\Matr{B}(E)\), such as
the Lyapunov exponent and the decay time of the classical Markov process, are
related to operator-growth velocities, entanglement-entropy growth, scrambling
times, and the exponential growth rates of OTOCs.
\end{enumerate}


\funding{Supported by a research grant from Magnus Konow in honour of his mother Olga Konow Rappaport.}


\clearpage
\appendix
\crefalias{section}{appendix}
\crefname{appendix}{Appendix}{Appendices}
\Crefname{appendix}{Appendix}{Appendices}
\makeatletter
\numberwithin{equation}{section}
\renewcommand{\theequation}{\thesection.\arabic{equation}}
\let\IOPorig@section\section
\renewcommand{\section}[1]{%
  \refstepcounter{section}%
  \addcontentsline{toc}{section}{Appendix~\thesection.\ #1}%
  \IOPorig@section*{Appendix~\thesection.\ #1}}
\newcommand{\restoreIOPsection}{\let\section\IOPorig@section}
\makeatother


\section{Estimate of the long-time entropy of quantum evolution}\label{Sec:Sinf}
Basis-dependent measures such as participation ratios and Shannon entropy have
long been used to characterize state spreading and randomness in quantum chaos;
see the recent review \cite{LakshminarayanZyczkowski2026} and references therein.
Here, our purpose is narrower: we estimate the long-time Shannon entropy
generated by a fixed quantum Poincar\'e map for a prescribed initial state.

We consider the energy-dependent unitary quantum evolution
\begin{equation}
    |\psi(t)\rangle = \Matr{U}^t|\psi(0)\rangle,
    \qquad t=0,1,2,\ldots,
\end{equation}
energy-dependent governed by a unitary quantum Poincar\'e map, \(\Matr{U}(E)\in\mathbb{C}^{D\times D}\), defined by \cref{Eq:MatrU} in \cref{Sec:PrelimQuantum}. 
The probabilities and Shannon
entropy are defined in the basis of the directed edges (the natural basis) $\{|k\rangle\}_{k=1}^{D}$:
\begin{equation}
    p_k(t)=|\langle k|\psi(t)\rangle|^2,
    \qquad
    S_{U}(t)=-\sum_{k=1}^{D}p_k(t)\ln p_k(t).
\end{equation}
All statements below refer to this fixed basis.
We use Latin indices for the directed-edge basis and Greek indices for the
eigenvectors and eigenvalues of \(\Matr{U}(E)\), so that:
\begin{equation}
\Matr{U}^{t} = \sum\limits_{\alpha=1}^{D}e^{\mathrm{i}\theta_{\alpha}t}|\alpha\rangle\langle\alpha|, \quad \theta_{\alpha}\in [0, 2\pi).
\end{equation}

We assume that, for \(t\gg1\), \(p_k(t)\) enters a long-time fluctuating regime
of the form \(p_k(t)=f_k+g_k(t)\), where \(f_k\) is its time average and
\(g_k(t)\) is quasiperiodic.
We make the corresponding assumption for the entropy and write:
\begin{equation}
S_{U}(t) \xrightarrow{t\gg 1} S_{U}^{(\infty)} + \delta S_{U}(t),  
\end{equation} 
where \(S_{U}^{(\infty)} = \langle S_{U}(t\gg 1)\rangle_{t}\) depends on the spectral properties of \(\Matr{U}(E)\) and the initial state preparation.

For an arbitrary initial state 
\[
|\psi(0)\rangle = \sum\limits_{i}a_{i}|i\rangle = \sum\limits_{i,\alpha}a_{i}|\alpha\rangle\langle\alpha|i\rangle, \quad \sum\limits_{i}|a_{i}|^{2} = 1,
\] 
\[
p_{k}(t) = \sum\limits_{\alpha, \beta} b_{k\alpha}b_{k\beta}^{\ast}e^{\mathrm{i}(\theta_{\alpha} - \theta_{\beta})t}, \quad b_{k\alpha} = \sum\limits_{i}a_{i}\langle k|\alpha\rangle\langle\alpha|i\rangle.
\]
We split the expression for \(p_{k}(t)\) into time-independent and time-dependent parts,  \(p_{k}(t) = f_{k} + g_{k}(t)\):
\begin{align}
&f_{k} = \sum\limits_{\alpha}|b_{k\alpha}|^{2} + \sum\limits_{\substack{\alpha\neq\beta\\
\theta_\alpha = \theta_\beta}}b_{k\alpha}b_{k\beta}^{\ast}\label{Eq:fk},\\
&g_{k}(t) = \sum\limits_{\substack{\alpha\neq\beta\\
\theta_\alpha \neq \theta_\beta}}b_{k\alpha}b_{k\beta}^{\ast}e^{\mathrm{i}(\theta_{\alpha} - \theta_{\beta})t}\label{Eq:gk},
\end{align}
where the second term in \cref{Eq:fk} accounts for degeneracies in the spectrum
of \(\Matr{U}(E)\).

Assume that \(|g_k(t)|\ll |f_k|\) for the times under consideration, \(t\gg 1\).  Expanding
each entropy contribution through second order gives
\begin{equation}\label{Eq:entexp}
-[f_k+g_k(t)]\ln[f_k+g_k(t)] =-f_k\ln f_k -g_k(t)[1+\ln f_k] 
      -\frac{g_k^2(t)}{2f_k}
       +\mathcal{O}\left(\frac{|g_k(t)|^3}{f_k^2}\right).
\end{equation}
If \(f_k=0\), then every spectral component at \(k\) vanishes and consequently
\(p_k(t)=g_k(t)=0\) for all \(t\); such an index is omitted from expressions
containing \(1/f_k\).
Introducing 
\[
G_{k} = \langle g_{k}^{2}(t)\rangle_{t} = \sum_{\substack{\alpha\neq\beta\\
\theta_{\alpha}\neq\theta_{\beta}}}\sum_{\substack{\gamma \neq \delta\\
\theta_{\gamma}\neq\theta_{\delta}}}b_{k\alpha}b_{k\beta}^{\ast}b_{k\gamma}b_{k\delta}^{\ast},  \quad \theta_{\alpha} - \theta_{\beta} + \theta_{\gamma} - \theta_{\delta} = 0\,(\mathrm{mod}\, 2\pi),
\]
we obtain:
\begin{equation}\label{Eq:SinfAppend}
S_{U}^{(\infty)} \approx  -\sum\limits_{k}\left(f_k\ln f_k + \frac{G_{k}}{2f_{k}}\right).
\end{equation}

\section{Approximate expressions of LE for Wigner-Dyson ensembles}
\label{SecAppend:LEapprox}

In this section we derive expressions for the ensemble-averaged LE,
\(\overline{\Lambda}(E)\), for the GOE and GUE in the limit
\(V=d+1\approx d\gg1\).
The notation \(\langle\cdot\rangle\) denotes the mean over all trajectories
generated by a single Hamiltonian matrix, whereas the overbar
\(\overline{\,\cdot\,}\) denotes the ensemble average.

GOE is an ensemble of real Hermitian matrices with real uncorrelated normally
distributed elements:
\begin{equation*}
\sbar{H^{}}{vw}=0,\quad
\sbar[1]{H^{2}}{v\neq w}=S_1^2=\frac{1}{4V},\quad
\sbar[1]{H^{2}}{vv}=2S_1^2=\frac{1}{2V},
\end{equation*}
while GUE is an ensemble of Hermitian matrices with complex uncorrelated
normally distributed elements
\(H_{vw}=a_{vw}+ib_{vw}\) (\(b_{vv}\equiv0\)):
\begin{equation*}
\sbar{a^{}}{vw}=\sbar{b^{}}{vw}=0,\quad
\sbar[1]{a^{2}}{v\neq w}
=
\sbar[1]{b^{2}}{v\neq w}
=
S_2^2=\frac{1}{8V},\quad
\sbar[1]{H^{2}}{vv}=2S_2^2=\frac{1}{4V}.
\end{equation*}

The expectation value of the absolute values of the off-diagonal matrix
elements,
\(h_{vw}=|H_{vw}|\), is
\begin{equation}
\xi=
\left\{
\begin{aligned}
&\xi_1=\frac{1}{\sqrt{2\pi V}}
\approx \frac{1}{\sqrt{2\pi d}}, &&\text{GOE},\\
&\xi_2=\frac{1}{4}\sqrt{\frac{\pi}{V}}
\approx \frac{1}{4}\sqrt{\frac{\pi}{d}}, &&\text{GUE}.
\end{aligned}
\right.
\end{equation}

By the Central Limit Theorem, the Gershgorin radius,
\(\Gamma_v=\sum_{w\neq v}h_{vw}\), approaches
\begin{equation}
\Gamma_v^2\xrightarrow{d\gg1}
\Gamma^2\left(1+\mathcal{O}(d^{-1})\right),
\qquad
\Gamma^2=\xi^2d^2\propto d.
\end{equation}
While higher-order corrections can be computed explicitly for both ensembles,
they contribute only \(o(d^{-1})\) terms to the LE. Accordingly, we restrict
the explicit derivation to the
\(\mathcal{O}(\ln d/d)\) and
\(\mathcal{O}(d^{-1})\) contributions.

As discussed in \cref{Sec:MeanLE} of the main text, the mean LE can be
computed using \cref{Eq:MeanLE-Th}:
\begin{equation}\label{Eq:MeanLE-Th-Apend}
\left\langle\Lambda(E)\right\rangle
=
-\frac{1}{D}
\sum\limits_{e,e'}
B_{e'e}(E)\ln B_{e'e}(E).
\end{equation}
To compute the ensemble average,
\(\overline{\Lambda}(E)\), we evaluate each term in the sum separately.
For each row of the matrix \(\Matr{B}\), there are \((d-1)\) elements
\((e,e')\) corresponding to transmission probabilities and one element
\((e,\hat e)\), where
\(\tau(e)=o(\hat e)\) and
\(\tau(\hat e)=o(e)\), corresponding to the reflection probability.

For GOE:
\begin{multline}\label{Eq:GOEstep1}
-\sbar{B_{e^{\prime}e}(E)\ln B_{e^{\prime}e}(E)}{}=\\
-\int\limits_{-\infty}^{+\infty}\frac{dH_{vw}}{\sqrt{2\pi S_{1}^{2}}}\int\limits_{-\infty}^{+\infty}\frac{dH_{vp}}{\sqrt{2\pi S_{1}^{2}}}\int\limits_{-\infty}^{+\infty}\frac{dH_{vv}}{\sqrt{4\pi S_{1}^{2}}}\exp\left\{-\frac{H_{vw}^{2} + H_{vp}^{2}}{2S_{1}^{2}} - \frac{H_{vv}^{2}}{4S_{1}^{2}}\right\}\times\\
\frac{4h_{vw}h_{vp}}{(H_{vv} - E)^{2} + \Gamma^{2}}\ln\left[\frac{4h_{vw}h_{vp}}{(H_{vv} - E)^{2} + \Gamma^{2}}\right] = \\
-\frac{d^{3}}{\pi^{3}\sqrt{2}}\int\limits_{-\infty}^{+\infty}dx_{v}\int\limits_{0}^{+\infty}dx_{w}\int\limits_{0}^{+\infty}dx_{p}
\exp\left\{-\frac{d^{2}}{4\pi}(x_{w}^{2} + x_{p}^{2}) - \frac{d^{2}}{2\pi}x_{v}^{2}\right\}\times\\
\frac{1}{1 + \left(x_{v} - E/\Gamma\right)^{2}}\left(-x_{w}x_{p}\ln\left[1 + \left(x_{v} - E/\Gamma\right)^{2}\right] + x_{p}x_{w}\ln x_{w} + x_{w}x_{p}\ln x_{p} \right),
\end{multline}
The integrals over \(x_{w}\), \(x_{p}\) can be evaluated exactly:
\[
-\int\limits_{0}^{+\infty}dx_{p}\int\limits_{0}^{+\infty}dx_{w}\exp\left\{-\frac{d^{2}}{4\pi}(x_{p}^{2} + x_{w}^{2})\right\}x_{w}x_{p} = 
-\frac{4\pi^{2}}{d^{4}}
\]
\[
\int\limits_{0}^{+\infty}dx_{p}\int\limits_{0}^{+\infty}dx_{w}\exp\left\{-\frac{d^{2}}{4\pi}(x_{p}^{2} + x_{w}^{2})\right\}\left(x_{w}x_{p}\ln x_{p} + x_{p}x_{w}\ln x_{w} \right) = 
-2\frac{2\pi}{d^{2}}\frac{\pi}{d^{2}}\left[\gamma + \ln\left(\frac{d^{2}}{4\pi}\right)\right], 
\]
where \(\gamma \approx 0.577\) is Euler--Mascheroni constant, so that \cref{Eq:GOEstep1} reads:
\begin{equation}\label{Eq:GOEStep3}
\frac{4}{\pi d\sqrt{2}}\int\limits_{-\infty}^{+\infty}dx_{v}e^{-\frac{d^{2}}{2\pi}x_{v}^{2}}
\left\{\frac{\ln\left[1 + \left(x_{v} - E/\Gamma\right)^{2}\right]}{1 + \left(x_{v} - E/\Gamma\right)^{2}} + \frac{\gamma + \ln\left(\frac{d^{2}}{4\pi}\right)}{1 + \left(x_{v} - E/\Gamma\right)^{2}}\right\}.
\end{equation}
A value of the integral is determined by the narrow vicinity of zero, \(\sim \sqrt{\pi}/d \ll 1\) as \(d \gg 1\). Thus, all the integrand functions besides the Gaussian exponent can be expanded in Taylor series with respect to \(x_{v}\). Taking in the account all \((d-1)\) transmission probabilities associated with \(v\)-th vertex, for transmission component of GOE ensemble average of LE one we obtain:
\begin{equation}\label{Eq:GOELambdaTr}
\sbar{\Lambda}{t}(E) = \frac{4(d-1)}{d^{2}}\frac{\ln\left[\frac{d^{2}\Delta^{2}}{4}\right]}{\Delta^2} + \frac{4(d-1)}{d^{2}}\frac{\gamma - \ln\pi}{\Delta^2} + \mathcal{O}\left(\frac{\ln (d)}{d^{3}}\right),
\end{equation}
where 
\[
\Delta^{2} = 1 + \left(\frac{E}{\Gamma}\right)^{2} = 1 + \left(\frac{E}{\xi d}\right)^{2}. 
\]
For the reflection term:
\begin{multline}\label{Eq:GOEreflect}
\sbar{\Lambda}{r}(E) = -\frac{2d^{2}}{\pi^{2}\sqrt{2}}\int\limits_{-\infty}^{+\infty}dx_{v}\int\limits_{0}^{+\infty}dx_{v}\exp\left\{-\frac{d^{2}}{\pi}x_{w}^{2} - \frac{d^{2}}{2\pi}x_{v}^{2}\right\}\times\\
\frac{\left(x_{v} - E/\Gamma\right)^{2} + (1 - 2x_{w})^{2}}{1 + (x_{v} - \eps)^{2}}\ln\left[\frac{\left(x_{v} - E/\Gamma\right)^{2} + (1 - 2x_{w})^{2}}{1 + \left(x_{v} - E/\Gamma\right)^{2}}\right] = \frac{4}{\Delta^2}\frac{1}{d} + \mathcal{O}\left(d^{-2}\right).
\end{multline}
Finally,
\begin{equation}\label{Eq:LambdaGOE-Apend}
\sbar{\Lambda}{\mathrm{GOE}}(E) = 
\frac{4}{d}\frac{\ln\left[\frac{d^{2}\Delta^2}{4}\right]}{\Delta^2} + \frac{4}{d}\frac{\gamma - \ln\pi + 1}{\Delta^2}  + o\left(d^{-1}\right),\quad \Delta^{2} = 1 + \left(\frac{E}{\xi_{1} d}\right)^{2}. 
\end{equation}

The derivation for GUE is analogous to that above, although it involves
integration over the real and imaginary parts of the off-diagonal matrix
elements. The final result is:
\begin{equation}\label{Eq:LambdaGUE-Apend}
\sbar{\Lambda}{\mathrm{GUE}}(E)
=
\frac{4}{d}
\frac{\ln\left[\frac{d^2\Delta^2}{4}\right]}{\Delta^2}
+
\frac{4}{d}
\frac{\gamma+\ln\pi-1}{\Delta^2}
+
o(d^{-1}),
\quad
\Delta^2=
1+\left(\frac{E}{\xi_2d}\right)^2 .
\end{equation}

It is instructive to compare
\cref{Eq:LambdaGOE-Apend,Eq:LambdaGUE-Apend}
with the mean LE of the \(d\)-regular graph,
\(\langle\Lambda_{\mathrm d}\rangle\), corresponding to
\(\Matr{H}^{(d)}=\xi_0\Matr{A}^{(\rm d)}\),
where
\(\xi_0=1/\sqrt{4(d-1)}\).
Using \cref{Eq:LambdaDregApprox} from the main text, we obtain:
\begin{equation}\label{Eq:LambdaDreg-Apend}
\sbar{\Lambda}{\mathrm d}(E)
=
\langle\Lambda_{\mathrm d}(E)\rangle
=
\frac{4}{d}
\frac{\ln\left[\frac{d^2\Delta^2}{4}\right]}{\Delta^2}
+
\frac{4}{d}
\frac{1}{\Delta^2}
+
o(d^{-1}),
\quad
\Delta^2=
1+\left(\frac{E}{\xi_0d}\right)^2 .
\end{equation}

For \(E=0\), the three expressions
\cref{Eq:LambdaGOE-Apend,Eq:LambdaGUE-Apend,Eq:LambdaDreg-Apend}
coincide at the leading asymptotic order,
\(\mathcal{O}(\ln d/d)\).
Their differences appear only at order
\(\mathcal{O}(d^{-1})\), giving
\[
\sbar{\Lambda}{\mathrm{GOE}}(0)
<
\sbar{\Lambda}{\mathrm{GUE}}(0)
<
\langle\Lambda_{\mathrm d}(0)\rangle .
\]
The inequality
\(\sbar{\Lambda}{\mathrm{GOE}}
<
\sbar{\Lambda}{\mathrm{GUE}}\)
follows from the relation \(e<\pi\).

\section{Mean Lyapunov Exponent of G\(\beta\)E\label{SecAppend:GbE-LE}}
We derive an estimate for
\(\sbar{\Lambda}{\mathrm{G\beta E}}\)
introduced in \cref{Sec:MLE-GbetaE}.  We employ the mean-field approximation, in which random variables are replaced
by their mean values.
The  distribution of the random variable \(b_{n}\) is given in \cref{Eq:ChiDistr} and two moments are given by:
\begin{equation}\label{Eq:ChiMoments}
\begin{aligned}
&\langle b_{n}\rangle = \sqrt{2}\frac{\Gamma\left(\frac{n\beta + 1}{2}\right)}{\Gamma\left(\frac{n\beta}{2}\right)}\xrightarrow{n\beta\gg 1}\sqrt{\beta n - \frac{1}{2}}\approx \sqrt{\beta n}\left(1 - \frac{1}{4\beta n}\right), \\
&\langle b_{n}^{2}\rangle = \beta n,
\end{aligned}
\end{equation}
thus
\begin{equation}\label{Eq:chi-distr-aux}
\begin{aligned}
&\langle (b_{n+1} - b_{n})^{2}\rangle = 1 + \mathcal{O}(1/n),\\ 
&\langle (b_{n+1} + b_{n})^{2}\rangle =  4\beta n + 2\beta -1 + \mathcal{O}(1/n),\\ 
&4\langle b_{n}b_{n+1}\rangle =  4\beta n + 2(\beta -1) + \mathcal{O}(1/n),
\end{aligned}
\end{equation}
while \(\beta = \mathcal{O}(1)\). The normally distributed random values \(a_{n}\) associated with the diagonal elements have the moments \(\langle a_{n} \rangle = 0\) and \(\langle a_{n}^{2} \rangle = 2\).

The classical Markov-Poincaré map, \(\Matr{B}\), is a product, \(\Matr{B}_{\mathrm{odd}}\cdot\Matr{B}_{\mathrm{even}}\), of the  block-diagonal matrices of size \(V-1\). Each block is a \(2\times2\)  matrix \(\matr{\pi}^{(n)}\):
\begin{equation}\label{Eq:GbEscatterrer}
\matr{\pi}^{(n)} = 
\begin{bmatrix}
p_{r}(n)&p_{t}(n)\\
p_{t}(n)&p_{r}(n)
\end{bmatrix}
\end{equation}
corresponding to an even or odd scatterer, with the exception of the two boundary scatterers. 
The transmission and reflection coefficients are given by
\begin{equation}\label{Eq:tnrnAppendix}
p_{t}(n, E) = \frac{4 b_{n} b_{n+1}}{\left(a_{n} - \sqrt{4\beta V}E\right)^{2} + \left(b_{n} + b_{n+1}\right)^{2}}, \quad 
p_{r}(n, E) = \frac{\left(a_{n} - \sqrt{4\beta V}E\right)^{2} + \left(b_{n} - b_{n+1}\right)^{2}}{\left(a_{n} - \sqrt{4\beta V}E\right)^{2} + \left(b_{n} + b_{n+1}\right)^{2}}.
\end{equation}

In the mean-field approximation one replaces the random coefficients
\(p_t\) and \(p_r\) by their ensemble-average values and writes:
\begin{equation}\label{Eq:Mean-tn}
\sbar{p}{t}(n, E) \approx \frac{4\mathbb{E}[ b_{n} b_{n+1}]}{\mathbb{E}\left[\left(a_{n} - \sqrt{4\beta V}E\right)^{2} + \left(b_{n} + b_{n+1}\right)^{2}\right]}
\xrightarrow{n\beta\gg 1} \frac{\frac{n}{V} + \frac{\beta - 1}{2\beta V}}{\frac{n}{V} + \frac{\beta - 1}{2\beta V} + E^{2} +\frac{3}{4\beta V}}
\end{equation}
\begin{equation}\label{Eq:Mean-rn}
\sbar{p}{r}(n, E) \approx \frac{\mathbb{E}\left[\left(a_{n} - \sqrt{4\beta V}E\right)^{2} + \left(b_{n} - b_{n+1}\right)^{2}\right]}{\mathbb{E}\left[\left(a_{n} - \sqrt{4\beta V}E\right)^{2} + \left(b_{n} + b_{n+1}\right)^{2}\right]}
\xrightarrow{n\beta\gg 1} \frac{E^{2} +\frac{3}{4\beta V}}{\frac{n}{V} + \frac{\beta - 1}{2\beta V} + E^{2} +\frac{3}{4\beta V}}
\end{equation}
The matrix \(\Matr{B}\) remains  bistochastic since \(\sbar{p}{r}(n, E) + \sbar{p}{t}(n, E) = 1\). The presence of the term \(3/(4\beta V)\) in the numerator of
\(\sbar{p}{r}\) is particularly convenient, since it dominates at
\(E=0\), ensuring that the reflection probability remains small but
nonzero.  

To estimate the LE, we introduce the continuous variable \(y=n/V\in[0,1]\), valid in the limit \(V\gg1\). We further introduce the shifted variable \(x = y + \theta\), \(\theta = (\beta - 1)/(2\beta V)\), and shifted energy, \(\widetilde{E}^{2} = E^{2} + 3/(4\beta V)\). Then \cref{Eq:Mean-tn,Eq:Mean-rn} become:
\begin{equation}\label{Eq:rxtx}
p_{r}(x, E) = \frac{x}{x + \widetilde{E}^{2}}, \quad p_{t}(x, E) = \frac{\widetilde{E}^{2}}{x + \widetilde{E}^{2}}.
\end{equation}

The bulk of \(\Matr{B}_{\mathrm{red}}\) consists products of the coupled blocks \(\matr{\pi}^{(n-1)}\cdot\matr{\pi}^{(n)}\), \(\matr{\pi}^{(n)}\cdot\matr{\pi}^{(n+1)}\). Replacing the elements of \cref{Eq:GbEscatterrer} by the continuous approximation, \cref{Eq:rxtx}, the mean LE can be written as:
\begin{equation}\label{Eq:LE-GbE-step1}
\sbar{\Lambda}{\mathrm{G\beta E}} = -\frac{1}{V-1}\sum\limits_{e^{\prime}e}B_{e^{\prime}e}\ln B_{e^{\prime}e}\approx \int\limits_{\theta}^{1+\theta}g(x;\widetilde{E})dx,
\end{equation}  
where 
\begin{equation}\label{Eq:LambdaIntegrand}
g(x; \widetilde{E}) = -\left[p_{r}^{2}\ln (p_{r}^{2}) + 2p_{r}p_{t}\ln(p_{r}p_{t}) + p_{t}^{2}\ln(p_{t}^{2})\right] = -2\left[p_{t}\ln p_{t} + (1 - p_{t})\ln(1 - p_{t})\right],
\end{equation}
a continuous counterpart of \(\Matr{G}|0)\), while \(B_{e^{\prime}e}\) already stand for the elements of \(\Matr{B}_{\mathrm{red}}\).
Here we use notation \(\overline{\Lambda}\) for the mean LE, since the mean-field approximation implicitly performs the ensemble averaging.
This expression relies on the approximation
\(p_{t}^{2}(n)\approx p_{t}(n)p_{t}(n\pm1)\approx p_{t}(n)p_{t}(n\pm2) \approx p_{t}^{2}(x)\) with analogous relations for the remaining products. The integration limits in \cref{Eq:LE-GbE-step1} can be simplified:
\[
\int\limits_{\theta}^{1+\theta}g(y;\widetilde{E})dy = \int\limits_{0}^{1}g(x;\widetilde{E})dx +  \left[\int\limits_{1}^{1+\theta}g(x;\widetilde{E})dx - \int\limits_{0}^{\theta}g(x;\widetilde{E})dx \right] = 
\int\limits_{0}^{1}g(x;\widetilde{E})dx + o\left(V^{-1}\right).
\]
Thus, the problem is reduced to evaluation of the integral:
\begin{equation}\label{Eq:IntLambdaGbE-Append}
\mathcal{I}(E) = \int\limits_{0}^{1}dx\left\{ \frac{x}{\widetilde{E}^{2} + x}\ln\left(\frac{x}{\widetilde{E}^{2} + x}\right) + \left(1 - \frac{x}{\widetilde{E}^{2} + x}\right)\ln\left(1 - \frac{x}{\widetilde{E}^{2} + x}\right)\right\},
\end{equation}
which can be evaluated exactly. Multiplication by \(-2\) then yields \cref{Eq:LE-analyticGbE} of the main text:
\[
\sbar{\Lambda}{\mathrm{G\beta E}} = 2\Big\{\widetilde{E}^{2}\Li_{2}(-1/\widetilde{E}^{2}) + (\widetilde{E}^{2} + 1)\ln(\widetilde{E}^{2} + 1) - \widetilde{E}^{2}\ln \widetilde{E}^{2}\left[\ln(\widetilde{E}^{2} + 1) - \ln \widetilde{E}^{2} + 1\right]\Big\},
\]
where \(\Li_{2}(\cdot)\) denotes the dilogarithm.
 The factor of 2 arises due to the reduction of the phase space, \(2(V-1)\longrightarrow V-1\), because by reducing the Markov-Poincaré map to \(\Matr{B}_{\mathrm{odd}}\cdot\Matr{B}_{\mathrm{even}}\) we effectively cover two directed edges in one step.

One can also deduce from \cref{Eq:LE-analyticGbE} that the maximum of the
mean LE is attained at
\(E_{\mathrm{max}}\approx\pm1/\sqrt{2}\), with
\[
\sbar{\Lambda}{\mathrm{max}}
\approx
-\sqrt{2}
+(3+\ln2)\ln3
-2\ln2
\approx1.2.
\]
\section{LE variance of G\(\beta\)E: the \(F\)-component\label{SecAppend:GbE-Fcomponent}}
The \(F\)-component of the LE variance is defined as:
\begin{equation}\label{Eq:VarF-Def-Appendix}
\left\langle\left(\delta\Lambda\right)^{2}\right\rangle_{F} = \frac{1}{D}\sum\limits_{e,e^{\prime}}B_{e^{\prime}e}\left[\ln B_{e^{\prime}e}\right]^2.
\end{equation}
Following the approach of \cref{SecAppend:GbE-LE}, we estimate it using the
same mean-field approximation:
\begin{equation}\label{Eq:VarF-Int}
\sbar{\left[\delta\Lambda(E)\right]^{2}}{F}\approx \int\limits_{0}^{1}dx\left\{p_{r}^2\bigl[\ln(p_{r}^2)\bigr]^2\;+\;2p_{r}p_{t}\bigl[\ln(p_{r}p_{t})\bigr]^2\;+\;p_{t}^2\bigl[\ln(p_{t}^2)\bigr]^2\right\},
\end{equation}
where \(p_r(x;E)=1-p_t(x;E)\) is defined in \cref{Eq:rxtx}. The integrand,
\(f[p_t(x),p_r(x)]\),
can be decomposed into two parts: one can be evaluated analytically,
whereas the other can be approximated using the results of
\cref{SecAppend:GbE-LE}. Indeed,
\begin{multline}
f[p_{t}(x), p_{r}(x)] = 4p_{r}^{2}[\ln p_{r}]^{2} + 4p_{t}^{2}[\ln p_{t}]^{2} + 2p_{r}p_{t}[\ln p_{r}]^{2} + 2p_{r}p_{t}[\ln p_{t}]^{2} + 4p_{r}p_{t}\ln p_{r}\ln p_{t} =\\
2p_{r}(2p_{r} + p_{t})[\ln p_{r}]^{2} + 2p_{t}(2p_{t} + p_{r})[\ln p_{t}]^{2} + 4p_{r}p_{t}\ln p_{r}\ln p_{t} =\\
 2\left(p_{r}[\ln p_{r}]^{2} + p_{t}[\ln p_{t}]^{2}\right) + 2\left(p_{r}\ln p_{r} + p_{t}\ln p_{t}\right)^{2}.
\end{multline}
The first contribution is precisely the integrand that would correspond to the
original Markov--Poincar\'e map \(\Matr{B}\). So, 
\[
\mathcal{I}_{1} = \int\limits_{0}^{1}dx\left(p_{r}[\ln p_{r}]^{2} + p_{t}[\ln p_{t}]^{2}\right)
\]
would be an \(F\)-component of the LE variance had we kept matrix \(\Matr{B}\) for the classical evolution, assuming that the presence of the eigenvalue \(-1\) in its spectrum is not
essential for reaching the steady state in most cases.This integral can be evaluated analytically similarly to \cref{Eq:IntLambdaGbE-Append}:
\begin{multline}
\mathcal{I}_{1}(E) = 
[\ln(1+\widetilde{E}^{2})]^{2} + \widetilde{E}^{2}\ln\left(\frac{\widetilde{E}^{2}}{1 + \widetilde{E}^{2}}\right)\ln(1+\widetilde{E}^{2})\left[\ln(1+\widetilde{E}^{2}) - 2\right] -\frac{\widetilde{E}^{2}}{3}\left[\ln\left(\frac{\widetilde{E}^{2}}{1 + \widetilde{E}^{2}}\right)\right]^{3}-\\
 2\widetilde{E}^{2}\left[\ln(1+\widetilde{E}^{2}) - 1\right]\Li_2\left(\frac{1}{1+\widetilde{E}^{2}}\right) - 2\widetilde{E}^{2} \Li_3\left(\frac{1}{1+\widetilde{E}^{2}}\right),
\end{multline}
where \(\widetilde{E}^{2} = E^{2} + 3/(4\beta V)\), and \(\Li_2(\cdot)\) and \(\Li_3(\cdot)\) denote the second- and third-order polylogarithms,
respectively.
The second component is too unwieldy for the exact analytical integration, but it can be approximated using the result of
\cref{Eq:IntLambdaGbE-Append}:
\[
\mathcal{I}_{2}(E) = 2\int\limits_{0}^{1}dx\left(p_{r}\ln p_{r} + p_{t}\ln p_{t}\right)^{2} = \frac{1}{2}\int\limits_{0}^{1}dx g^{2}(x; \widetilde{E})  \gtrsim
 \frac{1}{2}\left[\int\limits_{0}^{1}dx g(x; \widetilde{E})\right]^{2} = \frac{1}{2}\left[\overline{\Lambda}(E)\right]^{2},
\]
where \(g(x; E)\) is defined by \cref{Eq:LambdaIntegrand}. Finally, the \(F\)-component of the LE variance together with the centering
term is approximated by
\begin{equation}\label{Eq:FcomponentGbE}
\sbar{\left[\delta\Lambda(E)\right]^{2}}{F}
-
\left[\overline{\Lambda}(E)\right]^2
\approx
2\mathcal{I}_{1}(E)
-
\frac{1}{2}\left[\overline{\Lambda}(E)\right]^2,
\end{equation}
see \cref{Fig:GbE_Var}(c).

\section{Thermal Average Derivations\label{SecAppend:ThermalAverage}}
\subsection{The Lyapunov Exponent}\label{SecAppend:ThermalAvLE}

For \(V\gg1\), the thermal average can be expressed as
\begin{equation}\label{Eq:ThermalAverageDef}
\langle\Lambda(T)\rangle_{\mathrm{th}}
=
\frac{1}{\mathcal{Z}(T)}
\int\limits_{-\infty}^{+\infty} dE\,\rho(E)\Lambda(E)e^{-E/T},
\quad
\mathcal{Z}(T)
=
\int\limits_{-\infty}^{+\infty} dE\,\rho(E)e^{-E/T},
\end{equation}
where \(\rho(E)\) is the corresponding spectral density.
Here we assume that the LE has already been averaged over trajectories and the
ensemble, and denote it simply by \(\Lambda(E)\) to avoid introducing too many
averaging notations.
Since explicit expressions for the LE are already available for each model,
the thermal average can be computed straightforwardly using the corresponding
spectral density.

For the \(d\)-regular graph, one can use the Kesten--McKay distribution
\cite{McKay1981}:
\begin{equation}\label{Eq:McKay}
\rho_{\mathrm{d}}(E) = \left\{
\begin{aligned}
&\frac{d}{2\pi}\frac{\sqrt{4(d-1) - \lambda^{2}}}{d^{2} - \lambda^{2}}, &|\lambda|\leqslant 2\sqrt{d-1}, \\
&0,&\text{otherwise},
\end{aligned}
\right.,\quad \lambda = E/\xi.
\end{equation}
together with \cref{Eq:LambdaDreg} for
\(\Lambda_{\mathrm d}(E)\).
Then, for \(d=2\), one obtains:
\begin{equation}\label{Eq:LEthermalDreg2}
\langle\Lambda_{2}(T)\rangle_{\mathrm{th}} = \frac{a_0}{2} + 
\sum\limits_{m=1}^{\infty} a_{2m}\dfrac{I_{2m}(\widetilde{\beta}_{2})}{I_0(\widetilde{\beta}_{2})},\, \widetilde{\beta}_{2} = \frac{2\xi}{T},
\end{equation}
\[
a_{2m}=\frac{2}{\pi}\int_{0}^{\pi}
\Lambda(2\cos\theta)\cos(2m\theta)\,d\theta,
\]
where \(I_n\) is the modified Bessel function of the first kind \cite{AbramowitzStegun1964}.
For \(d>2\), the thermal average of the LE is given by:
\begin{equation}\label{Eq:LEthermalDregGen}
\langle\Lambda_{\mathrm{d}}(T)\rangle_{\mathrm{th}} = 
\frac{
\displaystyle
\sum\limits_{m=0}^{\infty}
c_{2m}^{\mathrm{(d)}}\,(2m+1)\,I_{2m+1}\big(\widetilde{\beta}_{\mathrm{d}}\big)
}{
\displaystyle
\sum_{m=0}^{\infty}
(d-1)^{-m}(2m+1)\,I_{2m+1}\big(\widetilde{\beta}_{\mathrm{d}}\big)
},
\quad\text{where}\quad
\widetilde{\beta}_{\mathrm{d}}=\frac{2\xi\sqrt{d-1}}{T},
\end{equation}
\[
c_{2m}^{\mathrm{(d)}}=\frac{2}{\pi}\int\limits_{-1}^{1}\frac{d(d-1)\sqrt{1-x^2}}{2\bigl(d^2-4(d-1)x^2\bigr)}U_{2m}^{ }(x)\Lambda_{\mathrm{d}}\bigl(2\sqrt{d-1}\,x\bigr)\,dx,
\]
where \(U_{n}(x)\) are the Chebyshev polynomials of the second kind \cite{AbramowitzStegun1964}. The choice of the normalization
\(\xi=1/\sqrt{4(d-1)}\)
implies the spectral support
\(|E|\le1\),
leading to the unified temperature scaling
\(\widetilde{\beta}_{\mathrm d}=1/T\).

For Wigner-Dyson and G\(\beta\)E ensembles, the spectral density is given by the Wigner semicircle distribution:
\begin{equation}\label{Eq:SemicircleSpectralDensity}
\rho(E) = \left\{
\begin{aligned}
&\frac{2}{\pi}\sqrt{1 - E^{2}},\, |E|\leqslant 1,\\
&0,\, |E|>1.
\end{aligned}
\right.
\end{equation}  
Thus, using \cref{Eq:LambdaGOEGUE} for the LE, the thermal average of the
Wigner--Dyson ensembles can be written as:
\begin{equation}\label{Eq:ThermalAverageGOEGUE}
\langle\Lambda_{\mathrm{WD}}(T)\rangle_{\mathrm{th}} \approx
\frac{4}{d}\left(2\ln\frac{d}{2}+A\right)
-
\frac{1}{\varkappa d^{2}}\left(2\ln\frac{d}{2}+A - 1\right)
\left[
1+3\frac{I_{3}(1/T)}{I_{1}(1/T)}
\right],
\end{equation}
where the constants are given by:
\[
\begin{aligned}
&\varkappa = 1/(2\pi),\, &A  = \gamma - \ln\pi + 1\quad& \text{for GOE},\\
&\varkappa = \pi/16,\, &A  = \gamma + \ln\pi - 1\quad& \text{for GUE}.
\end{aligned}
\]
It is worth noting that, for the GOE and GUE ensembles, the leading-order
asymptotics in \(d\) is independent of temperature, while
temperature-dependent corrections arise only at the next order.

For the G\(\beta\)E ensembles, we assume the same spectral density as in
\cref{Eq:SemicircleSpectralDensity}. The thermal average is then given by:
\begin{equation}\label{Eq:ThermalAverageGbE}
\langle\Lambda_{\mathrm{G\beta E}}(T)\rangle_{\mathrm{th}} = \sum\limits_{m=0}^{\infty}c_{2m}(2m + 1)\frac{I_{2m+1}(1/T)}{I_{1}(1/T)},
\end{equation}
\[
c_{2m}=\frac{2}{\pi}\int\limits_{-1}^{1}\Lambda_{\mathrm{G\beta E}}(E)\,U_{2m}(E)\,\sqrt{1-E^{2}}\,dE,
\]
where \(\Lambda_{\mathrm{G\beta E}}(E)\) is given by \cref{Eq:LE-analyticGbE}.

\subsection{Spectral gap}\label{Sec:ThermalAvSpectralGap}
For a \(d\)-regular graph, the computation of the thermal average differs
from that in \cref{SecAppend:ThermalAvLE}, since \(E\) cannot be treated as
a parameter independent of \(\alpha\) in \cref{Eq:DregGap}. 
Consider a single realization of the adjacency matrix
\(\Matr{A}^{\mathrm{(d)}}\), and the corresponding Hamiltonian
\(\Matr{H}^{\mathrm{(d)}}\). The energy takes values from the set
\(\{\alpha_i\xi\}\). The spectral gap is attained at \(E=\alpha_1\xi\). 
Therefore, the spectral gap of \(\Matr{B}\) is only a function of
\(\alpha_1\). 
Neglecting the distribution of \(\alpha_1\), which constitutes only a
finite-\(V\) effect beyond the scope of the present work, we obtain the
following temperature-independent estimate for the thermal average of the
spectral gap:
\begin{equation}\label{Eq:DregGap-vs-d}
\langle\varepsilon_{\mathrm{d}}\rangle_{\mathrm{th}} \gtrsim 1 - \frac{4\sqrt{d-1}}{4(d-1) + d^{2}} - \sqrt{\frac{16(d-1)}{[4(d-1) + d^{2}]^{2}} + 1 - \frac{4d}{4(d-1) +d^{2}}}, \quad d\geqslant 3.
\end{equation}

Applying the same formalism as in \cref{SecAppend:ThermalAvLE} to
\cref{Eq:GapWD}, one obtains the following thermal average of the spectral gap
for the GOE and GUE ensembles:
\begin{equation}\label{Eq:ThermalAverageGapWD}
\langle\varepsilon_{\mathrm{WD}}\rangle_{\mathrm{th}} = 
\left(1 - \frac{1}{4\varkappa d}\left[1 + 3\frac{I_{3}(1/T)}{I_{1}(1/T)}\right]\right)
\left\{
\begin{aligned}
&\frac{4\pi}{d^{3}},\,&\text{GOE},\\
&\frac{4\sqrt{2}}{d^{2}},\,&\text{GUE},
\end{aligned}
\right.
\end{equation}
where \(\varkappa\) is the same constant as in \cref{Eq:ThermalAverageGOEGUE}. 

Similarly, for the G\(\beta\)E spectral gap given by
\cref{Eq:GbEgapfull}, we obtain:
\begin{equation}\label{Eq:GbE-Gap-Thermal}
\langle\varepsilon_{\mathrm{G\beta E}}\rangle_{\mathrm{th}} = \sum\limits_{m=0}^{\infty}c_{2m}(2m + 1)\frac{I_{2m+1}(1/T)}{I_{1}(1/T)},
\end{equation}
where  the expansion coefficients are given by
\begin{equation}\label{Eq:C2mGbEgap}
c_{2m} = 2(-1)^{m}\frac{j_{1, 1}^{2}}{4V^{3/2}}\frac{(\sqrt{a_{\beta}/V+1} - \sqrt{a_{\beta}})^{2m}(a_{\beta}/V+ 1 - \sqrt{(a_{\beta}/V)(a_{\beta}/V+1)})}{\sqrt{a_{\beta}(a_{\beta}/V + 1)}}.
\end{equation}
Note the G\(\beta\)E gap dependence on the vertex dimension \(V\):
while, according to \cref{Eq:GbEgapfull}, the gap at fixed energy scales as
\(\propto V^{-\eta}\), \(1\leqslant\eta\leqslant2\), for any fixed temperature
its scaling becomes
\(\mathcal{O}(V^{-3/2})\).


\restoreIOPsection

\bibliographystyle{apsrev4-1}
\bibliography{QuantumChaosRefs}
\end{document}